\documentclass[%
preprint,
 amsmath,amssymb,
 aps,
]{revtex4-2}

\usepackage{graphicx}
\usepackage{dcolumn}
\usepackage{bm}

\begin{document}


\title{Vanadium~superconducting~microwave~resonators~on~silicon~wafers}
\author{Y. Fujita,$^{1}$$\footnote{E-mail: y.fujita@aist.go.jp}$  Y. Urade,$^{1}$ Y. Hibino,$^{2}$ M. Tsujimoto,$^{3}$ K. Inomata,$^{1}$ G. Fujii,$^{1}$ and W. Mizubayashi$^{1,4}$}
\affiliation{
$^{1}$Global Research and Development Center for Business by Quantum-AI technology (G-QuAT), National Institute of Advanced Industrial Science and Technology, 1-1-1 Umezono, Tsukuba, Ibaraki 305-8568, Japan
}
\affiliation{
$^{2}$Research Institute for Hybrid Functional Integration, National Institute of Advanced Industrial Science and Technology, 1-1-1 Umezono, Tsukuba, Ibaraki 305-8568, Japan
}
\affiliation{$^{3}$Core Electronics Technology Research Institute,National Institute of Advanced Industrial Science and Technology, 1-1-1 Higashi, Tsukuba, Ibaraki 305-3565, Japan
}
\affiliation{$^{4}$Semiconductor Frontier Research Center, National Institute of Advanced Industrial Science and Technology, 16-1 Onogawa, Tsukuba, Ibaraki 305-8569, Japan}




\date{\today}

\begin{abstract}
Understanding the correlation between material properties and microwave losses in superconducting films is a crucial subject for developing low-loss materials for quantum circuits.
We focus on vanadium (V) as a novel material for superconducting quantum devices and discuss loss in V films in relation to their structural properties.
Using a sputtering method, we grow four V-film structures on (001)-oriented Si wafers, employing Nb and Ta as the buffer and capping layer materials, respectively: Nb/V/Ta, Nb/V, V/Ta, and V.
X-ray diffraction and atomic force microscopy reveal that the V films grown on the Nb buffer layers have higher uniformity of lattice orientation and smaller grain size than that directly grown on the Si wafer.
Coplanar waveguide resonators are fabricated from the four V-film structures, and averaged photon number ($\langle n_{\rm ph} \rangle$) dependences of internal quality factor ($Q_{\rm int}$) are obtained by performing microwave measurements.
By analyzing the obtained $Q_{\rm int}$ vs $\langle n_{\rm ph} \rangle$, it is found that loss at the V surface is dominated by $\langle n_{\rm ph} \rangle$-independent non-two-level-system (non-TLS) losses, which can be mitigated by introducing the Ta capping layer. 
Furthermore, the V films on the Nb buffer layers exhibit lower $Q_{\rm int}$ in the $\langle n_{\rm ph} \rangle$ range from 10$^{0}$ to 10$^{6}$ and higher non-TLS loss than that directly grown on Si wafers, even though the former has higher lattice-orientation uniformity than the latter.
Origins of these trends might be relevant to V oxides, of which presence at surfaces and grain boundaries in bulk regions in the V resonators is suggested by energy dispersive X-ray spectroscopy and X-ray photoelectron spectroscopy, and/or V hydrides.

\end{abstract}

\maketitle



\section{\label{sec:intro} INTRODUCTION}

Superconducting qubits, which are superconducting solid-state devices for quantum information processing, are one of the leading candidate devices to build architecture of quantum processors \cite{Devoret_Science, Kjaergaard_Ann}. 
Toward large-scale integration of the superconducting qubits, it is required to lower error rates for qubit operation by enhancing their coherence time, which can be extended by eliminating energy-loss and dephasing factors inherent in materials for the qubits consisting of Josephson junctions and microwave resonators \cite{Sage_JAP, Oliver_MRS, McRae_Rev, Leon_Science, Siddiqi_NatRev}. 
To date, origins and mitigation of losses in quantum devices fabricated from a variety of superconducting materials, including aluminum (Al) \cite{Connell_Al_APL, Wang_Al_APL, Megrant_Al_APL, Richardson_Al_SST, Dunsworth_Al_APL, Grunhaupt_Al_PRL, Earnest_Al_SST, Un_Al_SciAdv}, niobium (Nb) \cite{Kumar_Nb_APL, Macha_Nb_APL, Burnett_Nb_SST, Gambetta_Nb_IEEE, Verjauw_Nb_PRAP, Altoe_Nb_PRXQ, Castanedo_Nb_AdvFM, Sung_Nb_PRMat}, tantalum (Ta) \cite{Place_Ta_NatCom, Wang_Ta_npjQ, Crowley_Ta_PRX, Urade_Ta_APLMat, Lozano_Ta_MatQT, Lozano_Ta_AdvSci}, and several nitrides \cite{Vissers_TiN_APL, Ohya_TiN_SST, Bruno_NbTiN_APL, deGraaf_NbN_NatCom, Terai_NbN_ComMat}, have been explored through development of material systems and process technologies.

In the development of materials for superconducting qubits, it is crucial to understand the correlation between intrinsic structural properties, namely, grain and lattice properties, and microwave loss in superconducting films for designing low-loss material systems. 
However, despite valuable efforts to address this issue through the fabrication and measurement of qubits and resonators, it remains poorly understood.
A prior work on the Nb-based transmon qubits showed that the coherence (loss) can be correlated with grain size, voids at grain boundaries, and oxygen (O) diffusion along the grain boundaries of the Nb films \cite{Premkumar_Nb_ComMat}.
This trend was supported by an investigation of losses in nucleated and non-nucleated epitaxial $\alpha$-Ta resonators \cite{Alegria_Ta_APL}.
By contrast, it was reported that there was no statistical difference in two-level-system (TLS) loss between Ta-based resonators fabricated from $\alpha$-Ta films with two different grain sizes \cite{Jones_Ta_JAP}.
With regard to lattice structures, superconducting films with ordered lattice structures are generally supposed to offer lower loss than disordered ones \cite{McRae_Rev}.
Quantum devices based on highly oriented or epitaxially grown films have demonstrated high performance, i.e., long coherence times in qubits \cite{Place_Ta_NatCom, Wang_Ta_npjQ, Terai_NbN_ComMat, Bland_Ta_Nature} and high quality factors in microwave resonators \cite{Megrant_Al_APL, Richardson_Al_SST, Earnest_Al_SST, Ohya_TiN_SST, Gao_TiN_PRMat}.
On the other hand, recent reports on the Ta-based resonators implied that enhancing degree of structural ordering and uniformity of lattice structures in the superconducting films could not necessarily offer dominant contribution to mitigate loss and presence of buffer (seed) layers could enhance loss even if it contributed to promote structural ordering \cite{Singer_Ta_QCE, Schijndel_Ta_PRAP, Marcaud_Ta_ComMat, Dhundhwal_Ta_APL}.
To deeply understand the correlation between structural properties and losses in superconducting films by overcoming this puzzling situation, it is necessary to obtain information on impacts of film structures on losses for an extensive range of superconducting materials comprehensively, including those of which loss properties are completely unknown. 

In line with this idea, we feature vanadium (V) as a novel superconducting material whose microwave loss properties at cryogenic temperatures remain underexplored. 
V is an element of group 5B, which is the same group as Nb and Ta. 
In the body-centered cubic (bcc) structure, the lattice constant of V (3.03 $\rm \AA$) is close to that of Nb (3.29 $\rm \AA$) and $\alpha$-Ta (3.30 $\rm \AA$), hence the consistency among the lattice structures of V, Nb, and Ta is expected to enable us to utilize them as buffer and/or capping layers for each other even in structurally ordered multilayers. 
Indeed, the superconducting diode based on the epitaxial \lbrack Nb/V/Ta\rbrack$_{n}$ superlattice structure was demonstrated, implying their good structural consistency \cite{Ando_JMSJ, Ando_Nature}. 

Here, we study impacts of structural properties on losses in V films.
Using a DC sputtering method, we fabricate four V-film structures with and without Nb buffer and/or Ta capping layers on (001)-oriented silicon (Si) wafers: Nb/V/Ta, Nb/V, V/Ta, and V.
X-ray diffraction (XRD) and atomic force microscopy (AFM) are performed to clarify how the Nb buffer layers affect crystallinity and grain size of the V films, and electrical transport properties of the V films are measured to confirm their superconductivity.
We fabricate superconducting resonators from the V-film structures and perform microwave measurements to obtain averaged photon number ($\langle n_{\rm ph} \rangle$) dependences of internal quality factor ($Q_{\rm int}$).
By analyzing the $Q_{\rm int}$ vs $\langle n_{\rm ph} \rangle$, we find that $\langle n_{\rm ph} \rangle$-independent non-TLS losses dominantly contribute to total loss at the V surfaces and are mitigated by introducing the Ta capping layer.
Furthermore, we find that the V films on the Nb buffer layers exhibit lower $Q_{\rm int}$ and higher non-TLS loss than those grown on the Si wafers.
Possible origins of these trends are discussed in relation to the V oxides and/or hydrides in the V resonators, considering structural ordering and grain size in the V films.

\section{\label{sec:results} RESULTS}

\subsection{\label{sec:results-1} V film growth}

We examined growth of V films on Si wafers, employing Nb and Ta as materials for buffer and capping layers, respectively. 
A 200-nm-thick V film was deposited on a 5-nm-thick Nb buffer layer formed on (001)-oriented Si wafers, which were cleaned using buffered hydrofluoric acid (BHF) for 1 minute prior to the depositions, by using a DC sputtering system at room temperature (RT).
A Ta capping layer with a thickness of 5 nm was sputtered on the top of the V layer.
To verify impacts of the presence of the Nb buffer layer on crystallinity of the V film and to identify the V surface morphology, we also prepared the samples without the Nb buffer layer, the Ta capping layer, and both of the Nb and Ta layers.
As a result, the samples with four different structures were prepared: Si/Nb(5 nm)/V(200 nm)/Ta(5 nm), Si/Nb(5 nm)/V(200 nm), Si/V(200 nm)/Ta(5 nm), and Si/V(200 nm).
The crystallography and surface morphology of the V films were characterized using XRD and AFM, respectively.
Detailed conditions for the film deposition and characterization are summarized in Appendix~\ref{app:A}. 

\begin{figure}
\includegraphics[width=8.5cm]{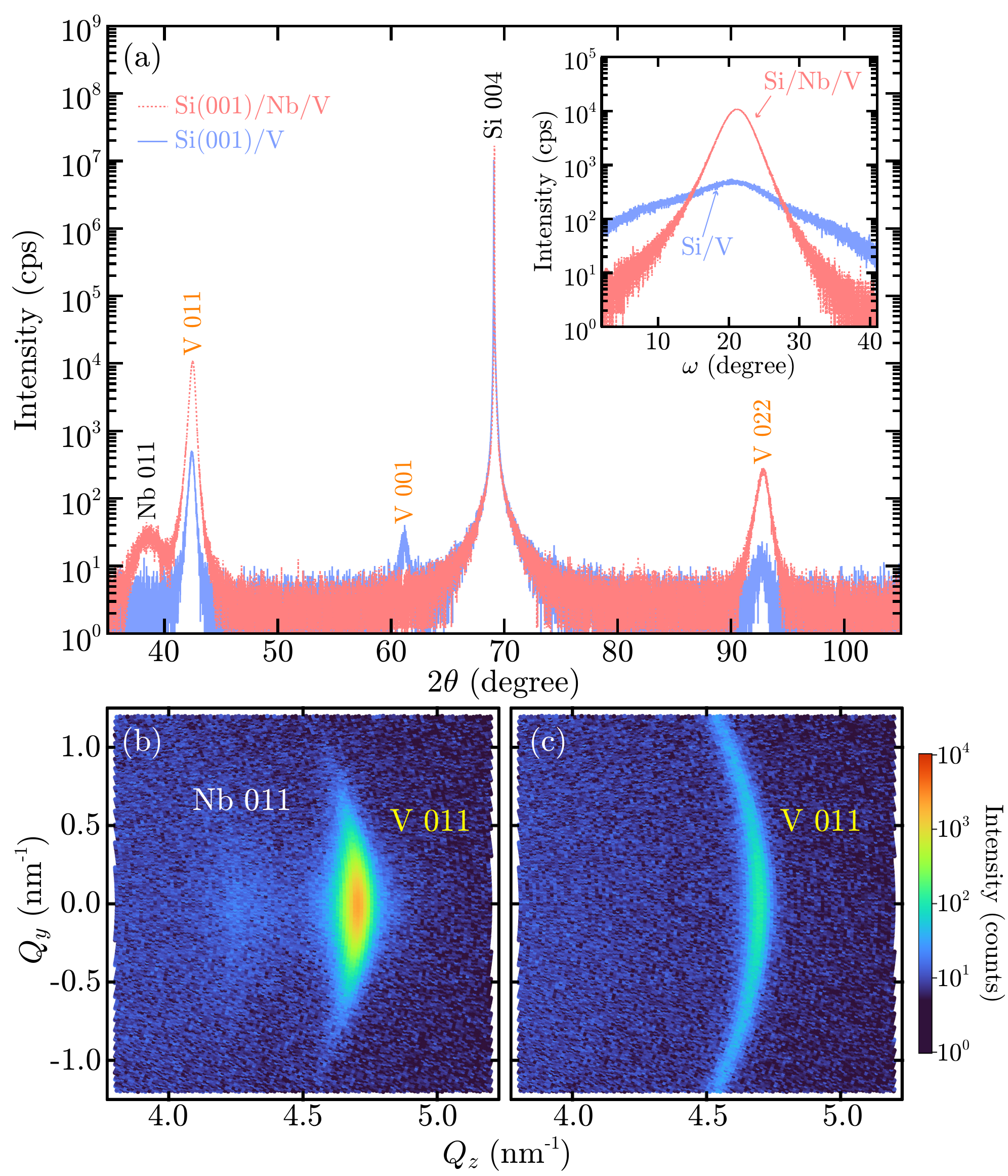}
\caption{\label{fig:1} (a) $\omega$-2$\theta$ XRD profiles for Si/Nb/V and Si/V samples obtained by setting the scattering vector to the out-of-plane ($\chi$ = 0$^\circ$) direction. The inset shows rocking curves of the observed V 011 peaks. RSMs around V 011 peaks for (b) Si/Nb/V and (c) Si/V structures.}
\end{figure}

Figure~\ref{fig:1}(a) shows the out-of-plane XRD profiles of $\omega$-2$\theta$ scans from the $\langle$001$\rangle$ directions of the samples with Si/Nb/V and Si/V structures.
The diffraction peaks of Nb 011, V 011, and V 022 were clearly observed in the Si/Nb/V structure, suggesting that the Nb and V layers were grown on the Si(001) substrate in an orientation such that (011)V$\parallel$(011)Nb$\parallel$(001)Si.
In addition to V 011 and V 022 peaks, a small V 001 peak was observed only in the Si/V structure, suggesting that the presence of the Nb buffer layer enhanced lattice coherence, i.e., uniformity of crystal orientation of V films along the perpendicular-to-plane direction.
This interpretation is supported by the rocking-curve measurements: the full width at half maximum (FWHM) of the V 011 peak is 13.24$^\circ$ for the Si/V sample, which is approximately three times larger than that for the Si/Nb/V sample (4.25$^\circ$), as shown in the inset of Fig.~\ref{fig:1}(a).
Note that the small V 001 peak was not always reproduced even when the film was deposited under the same setting conditions. 
This might be due to subtle fluctuations in target conditions and background and sputtering pressures for each deposition. 
Figures~\ref{fig:1}(b) and \ref{fig:1}(c) show the reciprocal space maps (RSMs) for the Si/Nb/V and Si/V structures, respectively. 
The Si/Nb/V structure clearly exhibits the V 011 peak with shapes different from that of the Si/V structure: the peak for the Si/V structure depicted an evident ring-like pattern, which has more significant broadening along the in-plane ($\chi$) direction than the Si/Nb/V structure.
This difference in peak shapes indicates that the Nb buffer layer also improves the in-plane lattice coherence of the V film.
Accordingly, the V film grown on the Nb buffer layer exhibits higher crystallographic-orientation uniformity than that grown directly on the Si wafer.

\begin{figure}
\includegraphics[width=8.5cm]{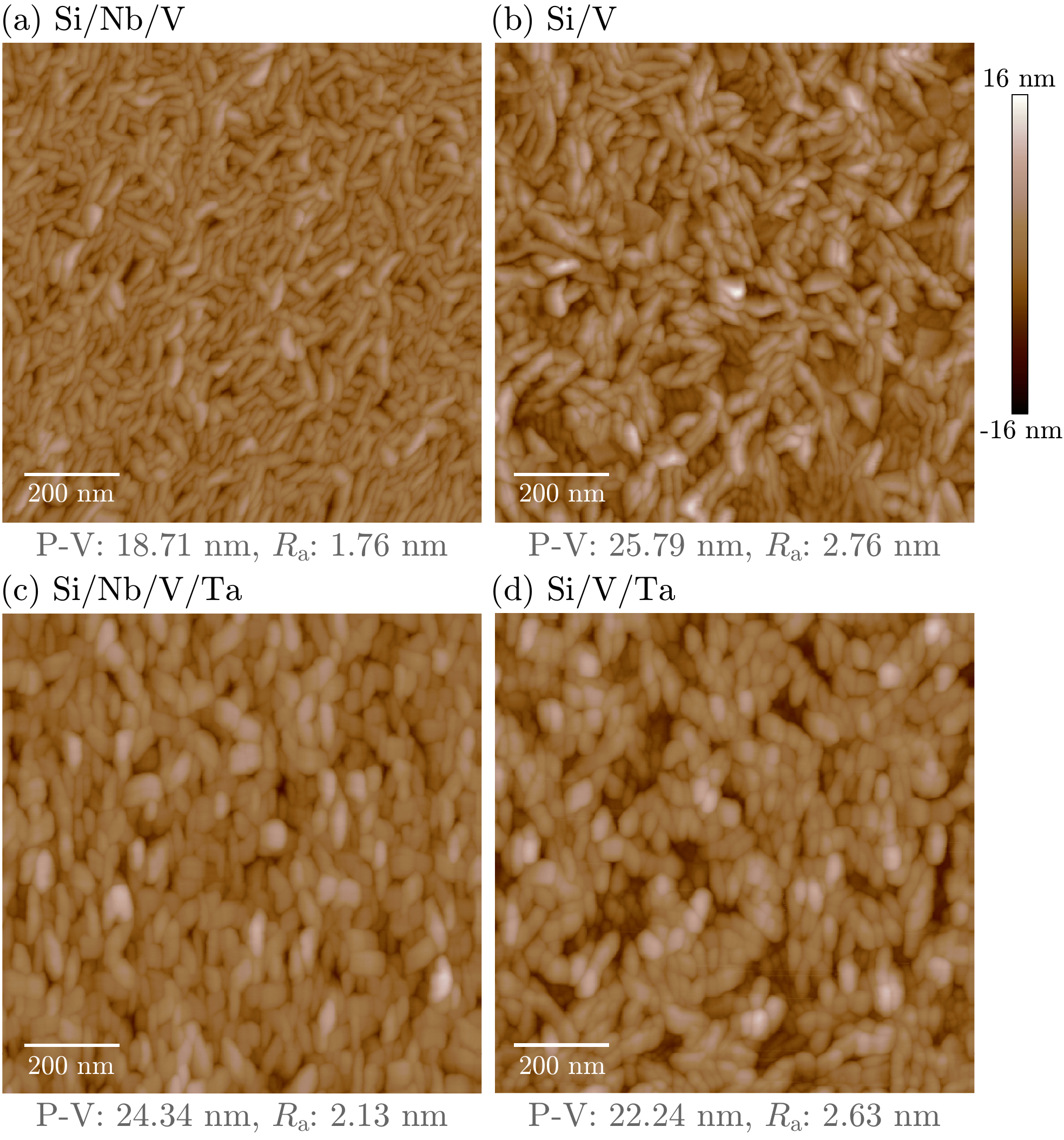}
\caption{\label{fig:2} AFM topographic images of the surfaces of (a) Si/Nb/V, (b) Si/V, (c) Si/Nb/V/Ta, and (d) Si/V/Ta structures. P-V and $R_{\rm a}$ are peak to valley and average surface roughness, respectively.}
\end{figure}

Figures~\ref{fig:2}(a) and \ref{fig:2}(b) show the AFM images of the surfaces of the V films grown on the Nb buffer layer and on the Si wafer, respectively.
Both of the V surfaces show ellipse and/or worm-like grain shapes, which were also observed on the surfaces of Nb and Ta films in previous reports \cite{Premkumar_Nb_ComMat, Jones_Ta_JAP, Marcaud_Ta_ComMat}, while numerous grains observed on the V surface of the Si/V structure appear larger than those of the Si/Nb/V structure.
Peak to valley (P-V) and average surface roughness ($R_{\rm a}$) for the Si/V structure were 25.79 nm and 2.76 nm, respectively, which were larger than those values for the V film grown on the Nb buffer layer of P-V = 18.71 nm and $R_{\rm a}$ = 1.76 nm, implying that the Nb buffer layer suppressed surface roughness of the V film.
The AFM images of the Ta surfaces of the Si/Nb/V/Ta and Si/V/Ta structures were shown in Figs.~\ref{fig:2}(c) and \ref{fig:2}(d), respectively.
Analysis of the AFM images and the corresponding P-V and $R_{\rm a}$ values indicates that the difference in grain size and surface morphology of the Ta capping layer between structures with and without the Nb buffer layer is smaller than that observed on the V surfaces.
Thus, surface roughness and grain size of the V film can be suppressed by the presence of the Nb buffer layer, though this trend was not fully preserved on the surface of the Ta capping layers formed on the V surfaces.

\subsection{\label{sec:results-2} Electrical transport properties}

To characterize the transport properties of the V films, we fabricated bar-shaped devices and measured the DC resistance ($R$) using a four-terminal method over a range of temperatures ($T$).
To assess the superconductivity of the thin Nb layer grown by the deposition method used in this study, we also prepared a small piece of a 5-nm-thick Nb film grown on a Si wafer and measured its $R$-$T$ characteristics separately.
In the measurements, temperature was swept from low to high.
Appendix~\ref{app:A} provides detailed specification of the samples, measurement conditions, and definitions of the superconducting transition temperature ($T_{\rm c}$) and normal-state resistance ($R_{\rm N}$), which are key parameters characterizing superconducting properties. 
\begin{figure}
\includegraphics[width=8.5cm]{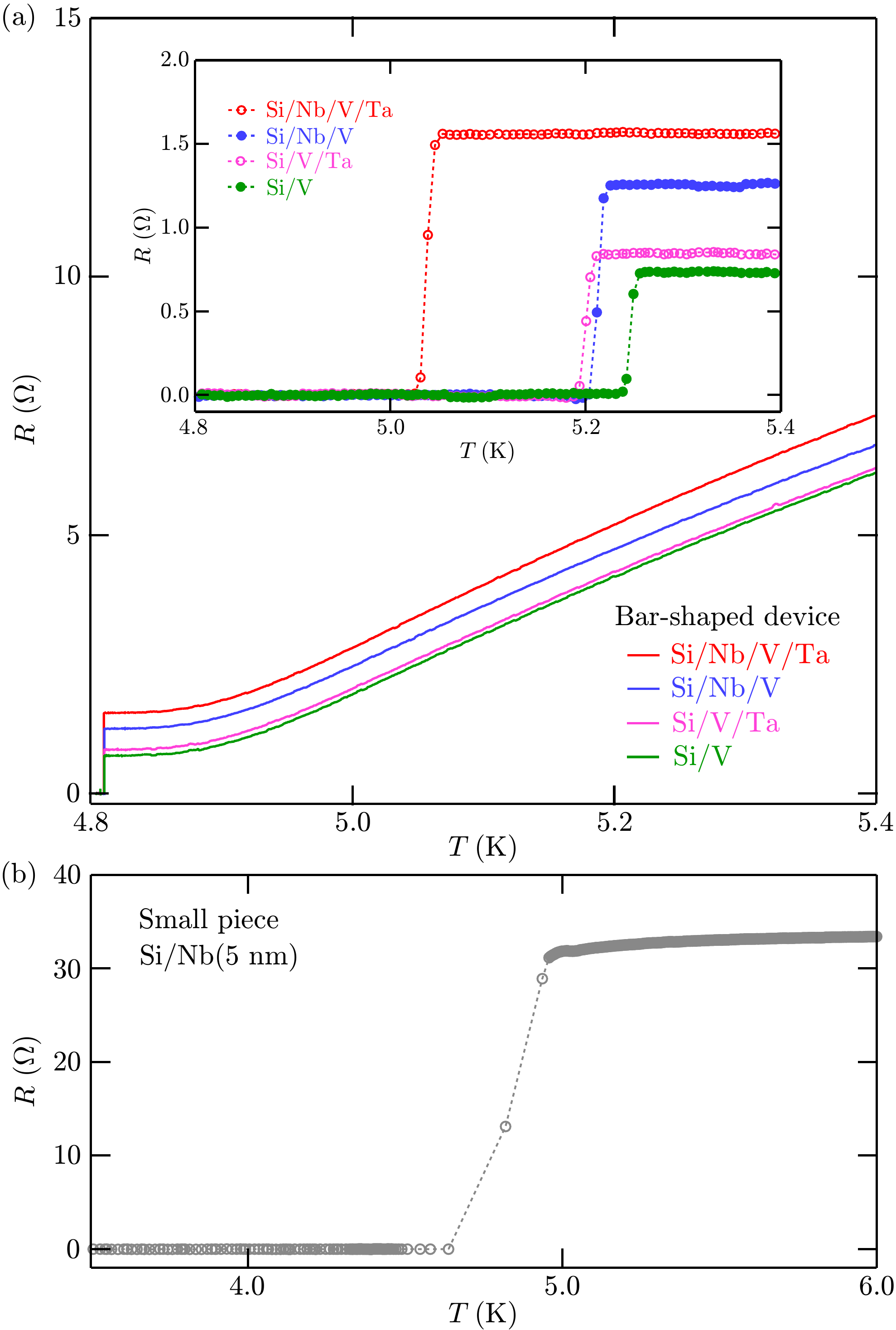}
\caption{\label{fig:3} (a) $T$ dependences of $R$ of bar-shaped devices of the four V-film structures. The inset shows the $R$-$T$ curves around the superconducting transition. (b) an $R$-$T$ curve of a small piece of the Si/Nb(5 nm) sample.}
\end{figure}

Figure~\ref{fig:3}(a) shows $R$ as functions of $T$ of the bar-shaped devices fabricated from the four different V-film structures in the range from 3 to 300 K.
All V-film structures exhibited a superconducting transition and typical metallic behavior, where $R$ increasing with increasing $T$ in the normal state.
$R$-$T$ curves around their superconducting transition are shown in the inset of Fig.~\ref{fig:3}(a).
All of the measured $T_{\rm c}$ values are close to the bulk values, consistent with previous reports that $T_{\rm c}$ of epitaxial V films is approximately 5.4 K \cite{Radebaugh_PR, Gutsche_TSF}.
Residual resistance ratio ($RRR$) of the V films is also obtained from the $R$-$T$ curves.
$RRR$ is defined as $R_{\rm 300 K}/R_{\rm N}$, where $R_{\rm 300 K}$ is the $R$ values at 300 K.
We summarized the obtained values of $R_{\rm 300 K}$, $R_{\rm N}$, and $RRR$ together with $T_{\rm c}$ in Table~\ref{tab:tab1}.
\begin{table}[t]
\caption{\label{tab:tab1} Parameters related to superconducting properties of the V films measured by the bar-shaped devices.}
\begin{ruledtabular}
\begin{tabular}{ccccc}
 Structure&$T_{\rm c}$ (K)&$R_{300 {\rm K}}$ ($\Omega$)&$R_{\rm N}$ ($\Omega$)&
 $RRR$\\
\hline
Si/Nb/V/Ta& 5.05 & 7.32 & 1.55 &4.71 \\
Si/Nb/V& 5.22 & 6.76 & 1.26 &5.37 \\
Si/V/Ta& 5.21 & 6.31 & 0.84 &7.50 \\
Si/V& 5.26 & 6.22 & 0.74 &8.44 \\
\end{tabular}
\end{ruledtabular}
\end{table}
$T_{\rm c}$ and $RRR$ of the Si/Nb/V structure are obtained to be 5.22 K and 5.37, respectively, which are lower than those of the Si/V structure (5.26 K and 8.44, respectively), indicating that the V films grown on the Nb buffer layers exhibit lower $T_{\rm c}$ and $RRR$ than those grown directly on the Si wafers.
Comparison of these values between the Si/Nb/V/Ta and Si/V/Ta structures revealed that this trend is observed independent of the presence of the Ta capping layer.
This reduction in $T_{\rm c}$ and $RRR$ by the Nb buffer layer may be owing to increase of the number of scattering events in the films caused by impurities in the layers, lattice defects at the Nb/V interface and/or smaller grain size, namely, higher grain boundary density in the V film grown on the Nb buffer layer than that directly grown on the Si wafer, as indicated by the AFM images displayed in Figs.~\ref{fig:2}(a) and \ref{fig:2}(b).
In addition, a proximity effect at the normal metal/superconductor interface \cite{Werthamer_PR, Gennes_RMP, Arrabal_JLTP, Weber_SST}  could account for the reduction in $T_{\rm c}$.
As shown in Fig.~\ref{fig:3}(b), the superconducting transition is clearly observed for the 5-nm-thick Nb film at $T_{\rm c}$ of approximately 4.94 K, which is clearly lower than that for the V-film structures without the Nb buffer layer. 
Note that the observed $T_{\rm c}$ of the 5-nm-thick Nb film is significantly lower than that of bulk Nb of 9.25 K \cite{Finnemore_PR}.
This is consistent with the well-documented thickness dependence of $T_{\rm c}$ of Nb films, where $T_{\rm c}$ decreased with reducing the thickness \cite{Mayadas_JAP, Wolf_JVST, Kodama_JAP, Minhaj_PRB}.
Based on this result, assuming that $T_{\rm c}$ of the thin Nb buffer layer is also lower than that of the adjacent V layer, the normal-state Nb/superconducting V interface can be formed in the V-film structures with the Nb buffer layers within the temperature range between specific $T_{\rm c}$ of the Nb buffer layer and the V layer.
The normal-state Nb/superconducting V interface can offer a proximity effect, suppressing $T_{\rm c}$ of the V layer.
We also found that the presence of the Ta capping layer suppressed $T_{\rm c}$ and $RRR$ from comparisons of those values between the Si/Nb/V/Ta and Si/Nb/V (Si/V/Ta and Si/V) structures.
Since $T_{\rm c}$ of the thin Ta capping layer should be lower than that of bulk Ta of approximately 4.4 K \cite{Read_APL}, which is expected to be much lower than specific $T_{\rm c}$ of the V layer, a proximity effect at the V/Ta interface might suppress $T_{\rm c}$ of the V layers adjacent to the Ta capping layers.

\subsection{\label{sec:results-3} Resonator fabrication}

Based on the structural and electrical characterization of the V films, we fabricated V-based superconducting resonators and performed microwave measurements.
Quarter-wavelength ($\lambda$/4) coplanar waveguide (CPW) resonators, designed identically to those in Ref.~\onlinecite{Urade_Ta_APLMat}, were fabricated from four types of V-film structures grown on 4-inch Si(001) wafers ($\rho$ $\ge$ 15 k$\Omega$cm) using conventional photolithography and reactive ion etching.
Detailed fabrication processes are described in Appendix~\ref{app:B}.
\begin{figure}
\includegraphics[width=8.5cm]{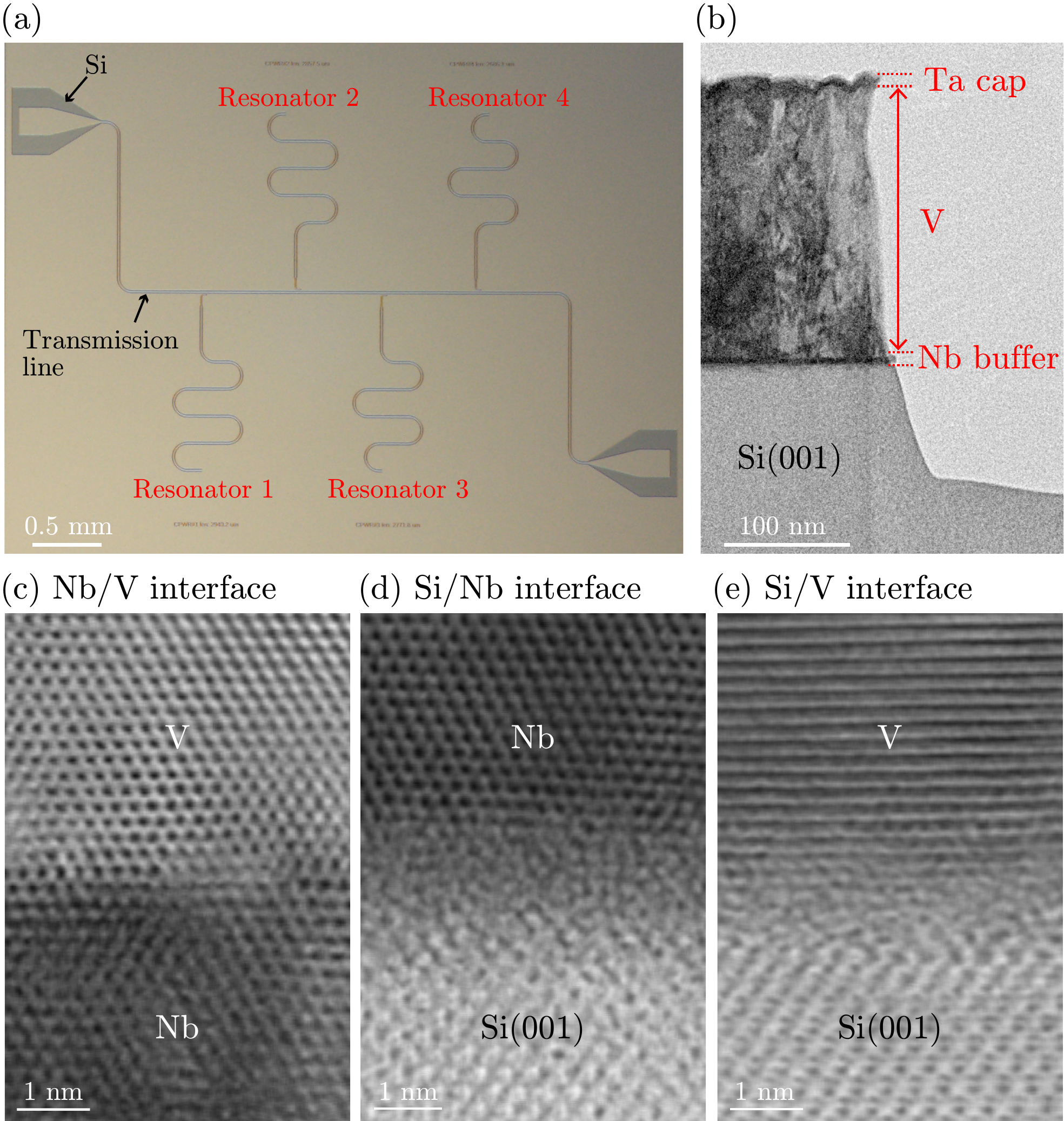}
\caption{\label{fig:4} (a) Optical micrograph of a CPW resonator chip with a Si/Nb/V/Ta structure. (b) Cross-sectional TEM image of the edge of the center conductor in a Si/Nb/V/Ta-structured resonator. High-resolution TEM images showing the (c) Nb/V and (d) Si/Nb interfaces in the Si/Nb/V/Ta-structured resonator, and (e) the Si/V interface in the Si/V/Ta-structured resonator.}
\end{figure}
The wafer samples were diced into 5 mm $\times$ 5 mm square chips, each containing four resonators labeled ``Resonator 1'' through ``Resonator 4'' in order of decreasing length, as shown in Fig.~\ref{fig:4}(a).
The four resonators were capacitively coupled to a single transmission line in the hanger mode \cite{McRae_Rev} and designed to have resonant frequencies ($f_{\rm c}$) around 10 -- 11 GHz.
The width of the center conductor and the gap between the center conductor and the ground plane were nominally 10 $\mu$m and 6 $\mu$m, respectively.
The structures of the resonators were analyzed using scanning transmission electron microscopy (TEM). 
The distribution of oxygen atoms was further examined by energy-dispersive X-ray spectroscopy (EDS), enabling estimation of both bulk and surface oxidation states.
\begin{figure}
\includegraphics[width=8.5cm]{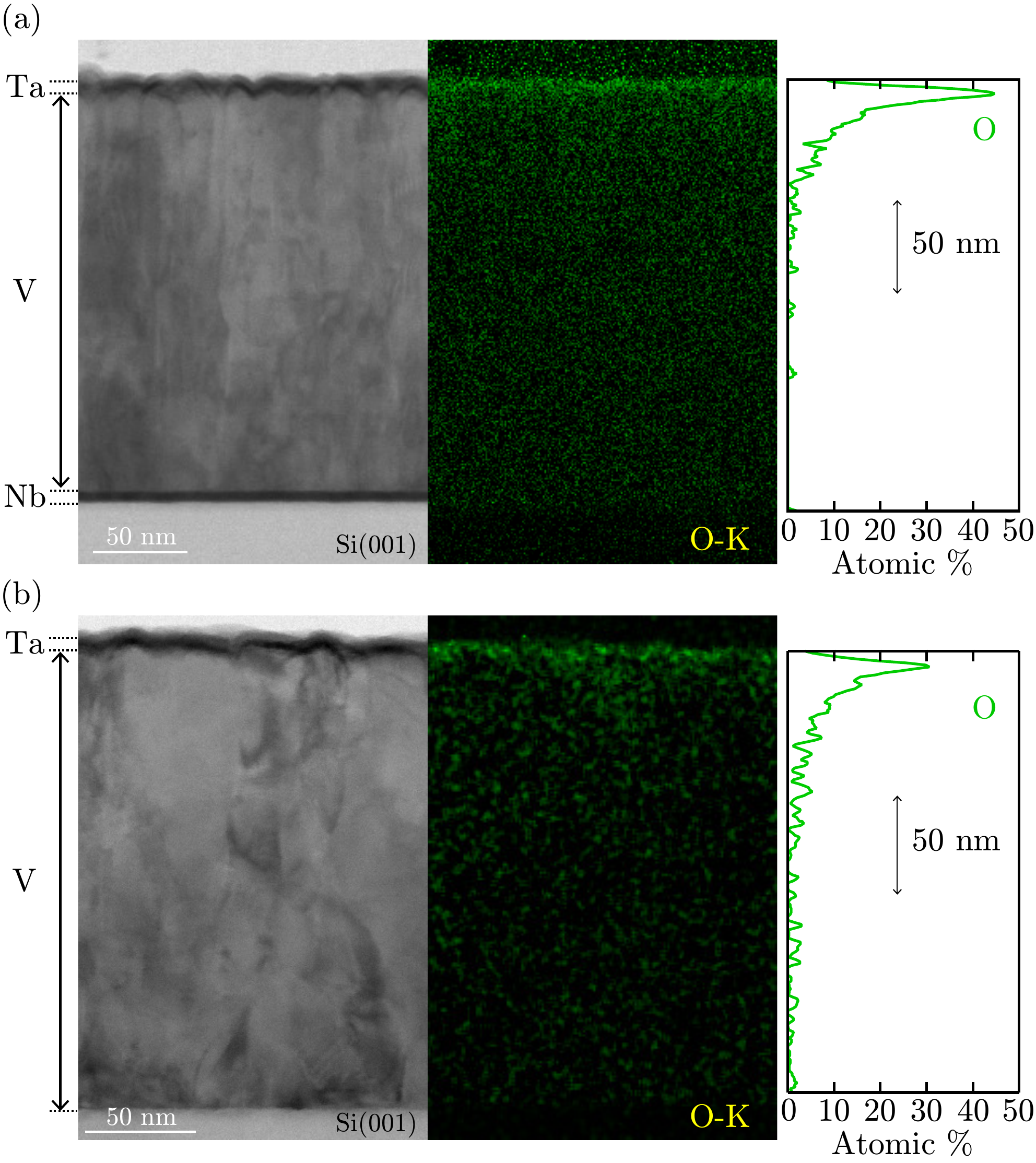}
\caption{\label{fig:5} Cross-sectional TEM images of center conductors (left) and EDS elemental maps (middle) and line concentration profiles in the depth direction of O (right) of (a) Si/Nb/V/Ta- and (b) Si/V/Ta-structured resonators.}
\end{figure}

Figure~\ref{fig:4}(b) shows a representative cross-sectional TEM image of an edge of the center conductor of the CPW resonator with the Si/Nb/V/Ta structure.
We confirmed that Si was etched to a depth of more than 100 nm beneath the Si/Nb interface with no significant atomic interdiffusion or interfacial reactions observed in the multilayered structure.
These features were also confirmed by TEM observations in the other fabricated structures, namely Si/Nb/V, Si/V/Ta, and Si/V (data not shown).
Figures~\ref{fig:4}(c) and \ref{fig:4}(d) show atomic-resolution TEM images of the Nb/V and Si/Nb interfaces in the resonator with the Si/Nb/V/Ta structure, respectively.
A continuous atomic arrangement was confirmed at the Nb/V interface, whereas a few-nm-thick light-contrast layer was observed between the Nb buffer layer and the Si wafer, as shown in Fig. 4(c), indicating the presence of a disordered layer composed of Nb and/or Si.
A similar layer, although less pronounced, was also observed at the Si/V interface in the Si/V/Ta-structured resonator as displayed in Fig.~\ref{fig:4}(e).
We infer that these disordered layers were owing to the large lattice mismatch between Si and the metals and/or atomic intermixing at the interfaces.

Figures~\ref{fig:5}(a) and \ref{fig:5}(b) show EDS elemental maps and oxygen concentration profiles on the center conductors of the Si/Nb/V/Ta and Si/V/Ta resonators, together with TEM images of the corresponding regions.
The oxygen elemental maps indicate surface-localized oxygen accumulation in both resonators, regardless of the presence of the Nb buffer layer.
By contrast, the oxygen concentration profiles reveal substantial oxygen distribution in the interior of the films, extending to depths greater than 50 nm below the surface despite the presence of Ta capping layers. 
This oxygen distribution is likely attributed to oxygen diffusion along near-surface grain boundaries, leading to the formation of V oxides at subsurface grain boundaries, as previously reported for Nb films \cite{Premkumar_Nb_ComMat}.

\subsection{\label{sec:results-4} Surface oxidation in resonators}

To further determine the detailed surface oxidation states of the resonators under the environment of microwave measurements, X-ray photoelectron spectroscopy (XPS) was performed for the resonator chips.
XPS spectra were observed in the regions of ground planes between Resonators 1 and 3 to analyze the metal surfaces, and from the Si-exposed gap region around the CPW transmission-line bonding pad to analyze the Si surface \lbrack see Fig.\ref{fig:4}(a)\rbrack.
Prior to microwave measurements, the chips were immersed in BHF solution for 1 min to remove native oxides from the exposed Si surfaces of the resonators, which are known to contribute to TLS defects, before being mounted in the dilution refrigerator system.
To reproduce the surface condition of the resonator chips during microwave measurements, the chips were loaded into the XPS chamber after approximately 3 hours of air exposure following the BHF treatment, corresponding to the typical time required for installation in the dilution refrigerator system.
Detailed conditions of the XPS measurements and analysis are summarized in Appendix~\ref{app:C}.

\begin{figure*}
\includegraphics[width=17cm]{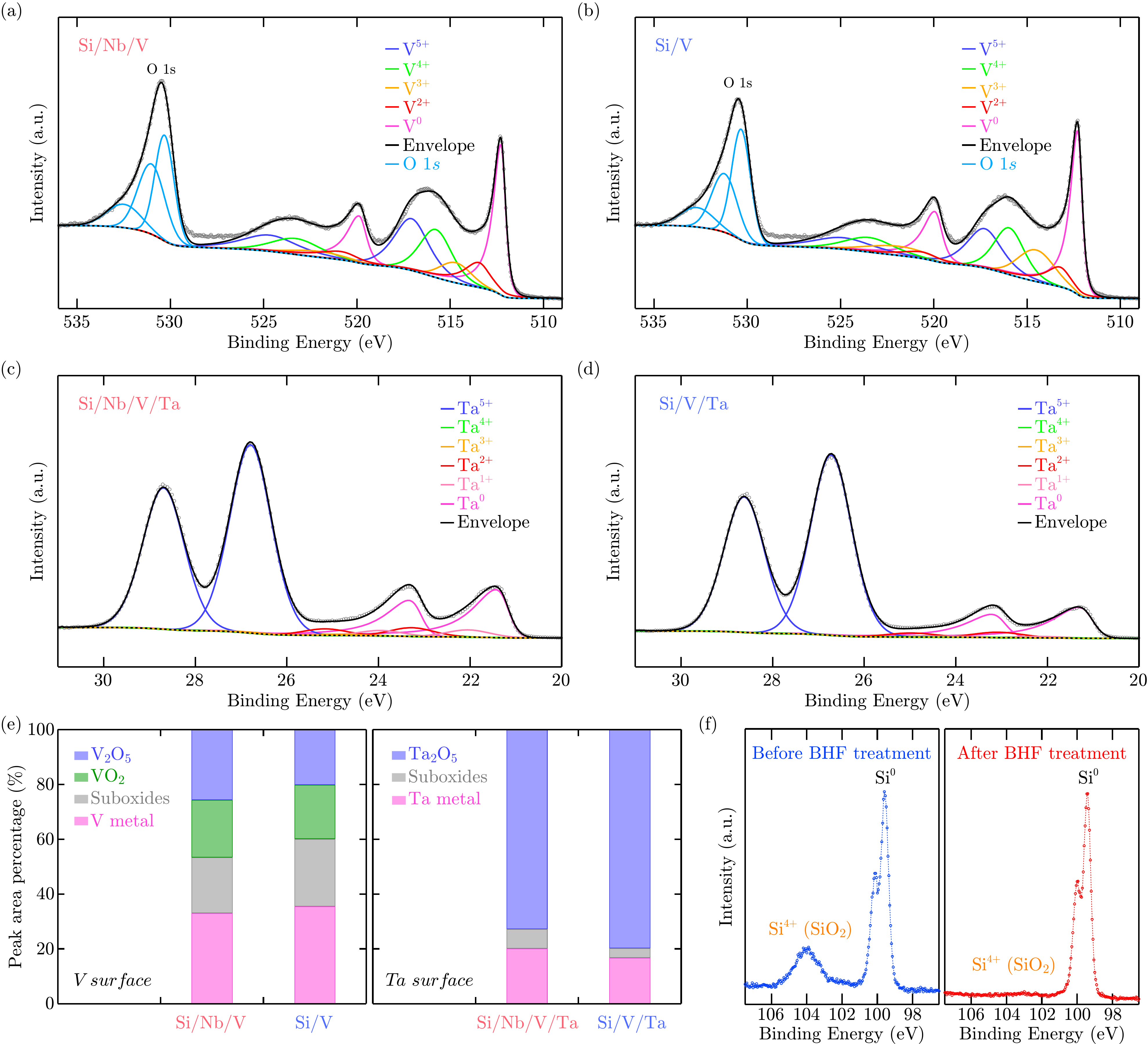}
\caption{\label{fig:6} V 2$p$ and O 1$s$ XPS spectra of (a) Si/Nb/V and (b) Si/V resonator chips. The O 1$s$ peaks were considered to determine the Shirley background (black dashed curves). Ta 4$f$ XPS spectra of (c) Si/Nb/V/Ta and (d) Si/V/Ta resonator chips. (e) Peak area percentage of surface oxides of V surfaces (left) and Ta surfaces (right) after the BHF treatment. (f) Si 2$p$ XPS spectra of a Si/Nb/V/Ta resonator chip before (left) and after (right) the BHF treatment.}
\end{figure*}

Figures~\ref{fig:6}(a) and \ref{fig:6}(b) show the V 2$p$ XPS peaks obtained from the V surfaces of Si/Nb/V and Si/V resonators, respectively. 
The observed V 2$p$ spectra were fitted with V$^{5+}$, V$^{4+}$, V$^{3+}$, V$^{2+}$, and V$^{0}$ components with spin-orbit splitting of V 2$p_{3/2}$ and V 2$p_{1/2}$ doublets. 
V$^{5+}$, V$^{4+}$, and V$^{0}$ components are assigned to V$_{2}$O$_{5}$, VO$_{2}$, and metallic V, respectively.
The remaining components, namely, V$^{3+}$ and V$^{2+}$ are attributed to the V suboxides, which may include V$_{2}$O$_{3}$ and VO.
We also measured the O 1$s$ signals with the V 2$p$ spectra to accurately determine the Shirley background \cite{Silversmit_JESRP, Biesinger_ASS}.
The components of each oxidation state were drawn by the lineshapes of Gaussian, Gaussian-Lorentzian product, or asymmetric-Lorentzian functions.
There were almost no qualitative difference in shapes of the observed spectra and the components of the oxidation states between the Si/Nb/V and Si/V resonators.
This qualitative tendency was also confirmed in the Ta 4$f$ XPS spectra for the Ta surfaces of Si/Nb/V/Ta and Si/V/Ta resonators, as shown in Figs.~\ref{fig:6}(c) and \ref{fig:6}(d), respectively.
As in the case of the V 2$p$ spectra, we extracted Ta oxidation states from the observed Ta 4$f$ XPS spectra by fitting the spectra of spin-orbit-split Ta 4$f_{7/2}$ and Ta 4$f_{5/2}$ doublets components corresponding to Ta$^{5+}$, Ta$^{4+}$, Ta$^{3+}$, Ta$^{2+}$, Ta$^{1+}$, and Ta$^{0}$.
The Ta$^{5+}$ and Ta$^{0}$ components are assigned to Ta$_{2}$O$_{5}$ and metallic Ta, respectively.
The remaining components indicate the states of suboxides. 
We quantitatively evaluated the relative contribution of each oxidation state on the V and Ta surfaces by estimating the peak-area ratio of the oxidation-state components from the XPS spectra, as shown in Fig.~\ref{fig:6}(e).
The results indicate that V$_{2}$O$_{5}$, VO$_{2}$, and suboxide components each contribute significantly to the observed V 2$p$ XPS spectra, suggesting that the V surfaces of both the Si/Nb/V and Si/V resonators exhibit complex oxidation states.
By contrast, for the Ta surfaces, Ta$_{2}$O$_{5}$, the most stable oxide of Ta, was the predominant component, contributing more than 70 \% of the total spectral weight for both the Si/Nb/V/Ta and Si/V/Ta resonators, while the other suboxide components exhibited substantially smaller contributions.
This trend is consistent with the previous XPS analyses for Ta surfaces reported in the literature \cite{Lozano_Ta_MatQT, Brumbach_JVSTA, McLellan_AdvSci}.
We also performed the same series of V 2$p$ and Ta 4$f$ XPS measurements prior to the BHF treatments and found that the metallic V contribution on the oxidation states of the V surfaces increased, whereas the Ta surfaces were hardly affected by the BHF treatments (see Appendix~\ref{app:C}).

Figure~\ref{fig:6}(f) shows the representative Si 2$p$ XPS spectra observed for the Si/Nb/V/Ta resonator chip before and after the BHF treatment.
The Si$^{4+}$ 2$p$ peak associated with SiO$_{2}$ definitely disappeared after the BHF treatment, confirming that the treatment effectively removed the native Si oxide, which is a known source of TLS defects.

\subsection{\label{sec:results-5} Microwave characterization}

After undergoing the BHF treatment described above, the resonator chips were placed in light-tight sample packages made of gold-plated oxygen-free copper.
The chips were connected to printed circuit boards via Al bonding wires. 
The sample package and circuit board were of the same design as those used in Ref.~\onlinecite{Urade_Ta_APLMat}.
The packages were equipped with magnetic shields and mounted on the mixing-chamber stage of a dilution refrigerator. 
The complex transmission coefficient ($S_{21}$) was measured 10 times at each measurement point using a vector network analyzer (VNA), while sweeping $\langle n_{\rm ph} \rangle$ estimated from the probe power ($P$) as $\langle n_{\rm ph} \rangle = \frac{2Q^{2}P}{\hbar\omega_{\rm c}^{2}Q_{\rm ext}}$, where $Q$ is the total (loaded) quality factor, $Q_{\rm ext}$ is the external (coupling) quality factor, $\hbar$ is the reduced Planck constant, and $\omega_{\rm c}$ is the angular resonance frequency \cite{Bruno_APL}.
The temperature of the mixing-chamber stage was maintained approximately 10 mK during microwave measurements.
See Appendix~\ref{app:D} for the details of the measurement setup and fitting procedures.
\begin{figure}
\includegraphics[width=8.5cm]{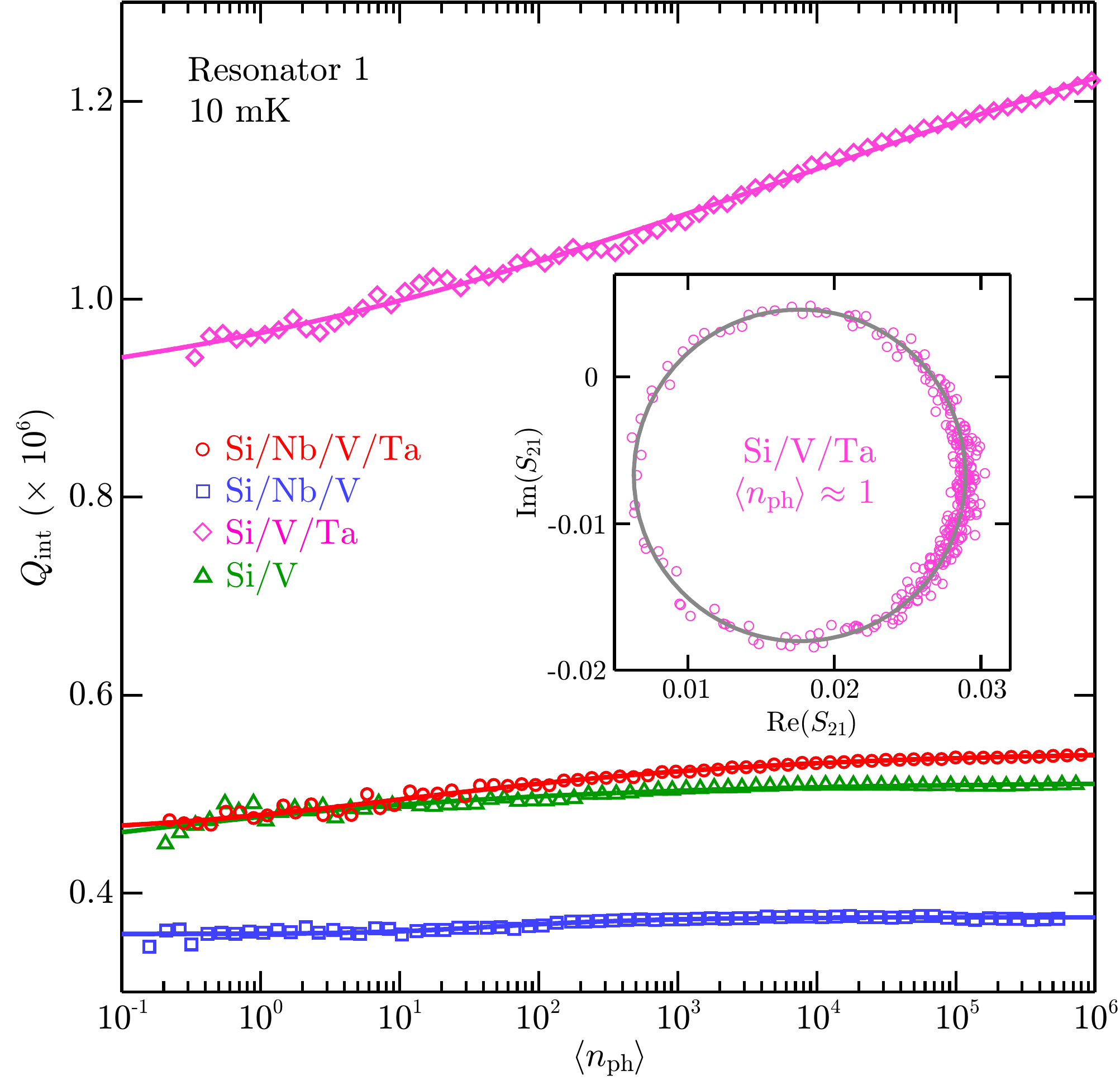}
\caption{\label{fig:7} $\langle n_{\rm ph} \rangle$ dependences of $Q_{\rm int}$ for Resonator 1 with Si/Nb/V/Ta (red circle), Si/Nb/V (blue square), Si/V/Ta (magenta diamond), and Si/V (green triangle) structures. The data points represent the median values of repeated measurements. Solid curves indicate fits to Eq.~(\ref{eq:1}). The inset shows a representative $S_{21}$ response in the complex plane at $\langle n_{\rm ph} \rangle \approx 1$ for the Si/V/Ta resonator, where the data and fitting curve used to extract $Q_{\rm int}$ are shown by magenta circles and a grey solid curve, respectively.}
\end{figure}

Figure~\ref{fig:7} shows $Q_{\rm int}$ as a function of $\langle n_{\rm ph} \rangle$ for Resonator 1 of the four structural types.
$Q_{\rm int}$ values were obtained by fitting the measured $S_{21}$ spectra with a theoretical model \cite{Khalil_JAP, Geerlings_APL}, even at the single-photon level ($\langle n_{\rm ph} \rangle \approx 1$), as representatively shown in the inset of Fig.~\ref{fig:7}.
We observed significantly weaker $\langle n_{\rm ph} \rangle$ dependences of $Q_{\rm int}$ than previously reported ones for $\alpha$-Ta and Nb resonators with the same design as the resonators in this study, in which $Q_{\rm int}$ at $\langle n_{\rm ph} \rangle \approx 10^{6}$ was more than an order of magnitude higher than that at $\langle n_{\rm ph} \rangle \approx 1$ \cite{Urade_Ta_APLMat}. 
This indicates that contribution of TLS-related loss to total loss, which decreases with increasing $\langle n_{\rm ph} \rangle$, is smaller in the V films than in $\alpha$-Ta and Nb films.
By comparing $Q_{\rm int}$ vs $\langle n_{\rm ph} \rangle$ for the Si/Nb/V/Ta (Si/V/Ta) with that for the Si/Nb/V (Si/V) resonators, we found that the resonators with the Ta capping layer exhibited a stronger $\langle n_{\rm ph} \rangle$ dependence of $Q_{\rm int}$, characterized by enhancement and saturation of $Q_{\rm int}$ with increasing $\langle n_{\rm ph} \rangle$, as well as higher $Q_{\rm int}$ over the entire $\langle n_{\rm ph} \rangle$ measurement range, compared with those without the Ta capping layer.
These results indicate that introducing the Ta capping layer mitigates $\langle n_{\rm ph} \rangle$-independent non-TLS losses, which otherwise strongly suppresses $Q_{\rm int}$ over a wide $\langle n_{\rm ph} \rangle$ range, while enhancing the relative contribution of TLS loss associated with dielectric Ta$_{2}$O$_{5}$ at the Ta surfaces in the V-based resonators.
Furthermore, comparison of $Q_{\rm int}$ vs $\langle n_{\rm ph} \rangle$ between the Si/Nb/V/Ta (Si/Nb/V) and Si/V/Ta (Si/V) resonators shows that the presence of the Nb buffer layer degrades $Q_{\rm int}$ over the entire range of $\langle n_{\rm ph} \rangle$, despite improving the crystallographic orientation uniformity of the V films, as suggested by the XRD results in Fig.~\ref{fig:1}.
This implies that improving structural ordering of the V films alone does not mitigate losses in the V-based resonators.
These trends were also observed for Resonator 2-4 (see Appendix~\ref{app:D}).

\begin{figure}
\includegraphics[width=8.5cm]{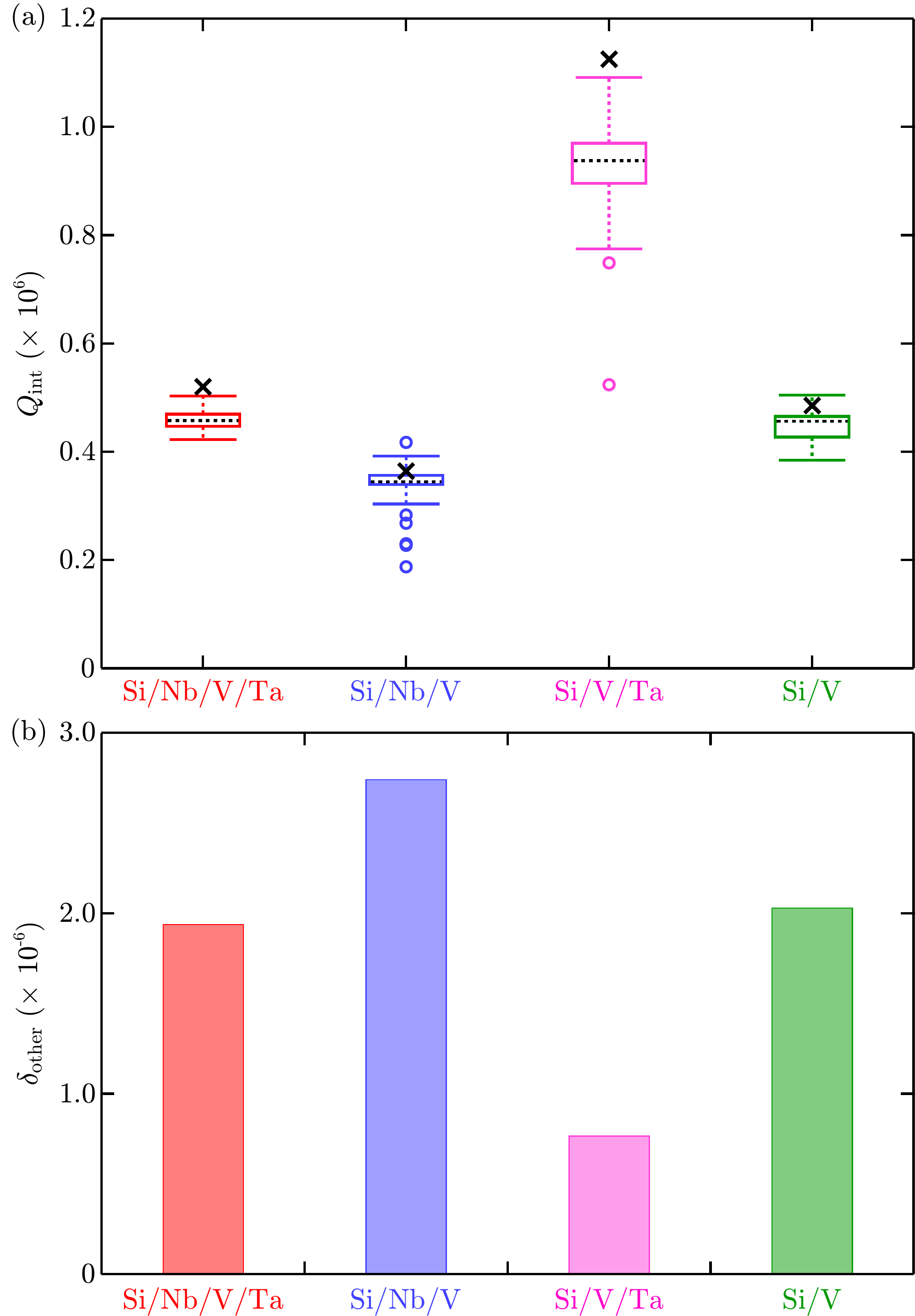}
\caption{\label{fig:8} (a) Box-and-whisker plots of $Q_{\rm int}$ at $\langle n_{\rm ph} \rangle \approx 1$ for resonators with the four V-film structures. The datasets were obtained from numerical fitting results of repeated $S_{21}$ measurements for all resonators on each chip. Open circles and black dashed lines indicate outliers and medians, respectively. The upper (lower) whisker denotes the highest (lowest) data point within $\lbrack {\rm Q}_{1} - 1.5 \times {\rm IQR},  {\rm Q}_{3} + 1.5 \times {\rm IQR} \rbrack$, where ${\rm Q}_{1}$ and ${\rm Q}_{3}$ are lower and upper quartiles, respectively, and IQR is the interquartile range defined as ${\rm Q}_{3} - {\rm Q}_{1}$. Black crosses indicate the median $Q_{\rm int}$ values at large photon numbers ($\langle n_{\rm ph} \rangle \approx 10^{5}$). (b) $\delta_{\rm other}$ values averaged over all resonators on each chip, obtained by fitting $Q_{\rm int}$ vs $\langle n_{\rm ph} \rangle$ with Eq.~(\ref{eq:1}). The fitting uncertainty is less than 3\%.}
\end{figure}

We statistically analyzed the observed trends in $Q_{\rm int}$ vs $\langle n_{\rm ph} \rangle$ over the full range of the microwave measurements in this study.
As shown in Fig.~\ref{fig:8}(a), $Q_{\rm int}$ at $\langle n_{\rm ph} \rangle \approx 1$ is summarized as box-and-whisker plots, including all the $Q_{\rm int}$ values extracted from repeated $S_{21}$ measurements of the four resonators on each chip.
The median $Q_{\rm int}$ values at $\langle n_{\rm ph} \rangle \approx 1$ for the Si/Nb/V/Ta, Si/Nb/V, Si/V/Ta, and Si/V resonators were 4.58 $\times$ 10$^{5}$, 3.45 $\times$ 10$^{5}$, 9.38 $\times$ 10$^{5}$, and 4.56 $\times$ 10$^{5}$, respectively.
Comparison of $Q_{\rm int}$ at $\langle n_{\rm ph} \rangle \approx 1$ between the Si/Nb/V/Ta (Si/V/Ta) and Si/Nb/V (Si/V) resonators shows that introducing the Ta capping layer clearly enhances it.
Likewise, comparison between the Si/Nb/V/Ta (Si/Nb/V) and Si/V/Ta (Si/V) resonators confirms that removing the Nb buffer layer reduces loss at $\langle n_{\rm ph} \rangle \approx 1$.
The same ordering of median $Q_{\rm int}$ was also observed at large photon number ($\langle n_{\rm ph} \rangle \approx 10^{5}$), as indicated by the black crosses in Fig.~\ref{fig:8}(a).
The increase in $Q_{\rm int}$ upon increasing $\langle n_{\rm ph} \rangle$ from approximately 1 to $10^{5}$ was estimated to be 0.62 $\times$ 10$^{5}$ (1.87 $\times$ 10$^{5}$) for the Si/Nb/V/Ta (Si/V/Ta) resonators, substantially larger than the corresponding value of 0.19 $\times$ 10$^{5}$ (0.29 $\times$ 10$^{5}$) for the Si/Nb/V (Si/V) resonators.
This suggests that the presence of the Ta capping layer strengthens the $\langle n_{\rm ph} \rangle$ dependence of $Q_{\rm int}$.
Therefore, the trends inferred from Fig.~\ref{fig:7} are consistently supported by the statistical analysis across all resonators and repeated measurements.

To clarify the origins of loss in the V-based resonators, we analyzed the observed $\langle n_{\rm ph} \rangle$ dependences of $Q_{\rm int}$ by fitting the $Q_{\rm int}$ vs $\langle n_{\rm ph} \rangle$ data with the TLS loss model as follows \cite{Crowley_Ta_PRX, Wang_APL}:
\begin{equation}
\label{eq:1}
\delta \left(\left\langle n_{\rm ph} \right\rangle\right) = \frac{F\delta_{\rm TLS, 0}}{{\sqrt{1 + \left(\frac{\left\langle n_{\rm ph} \right\rangle}{n_{\rm sat}}\right)^{\beta_{2}}}}} + \delta_{\rm other},
\end{equation}
where $\delta = Q_{\rm int}^{-1}$, $F$ is the participation ratio of electric energy in the region where TLS exists, $\delta_{\rm TLS, 0}$ is the intrinsic TLS loss in the zero-photon and zero-temperature limits, $n_{\rm sat}$ is the saturation photon number, $\beta_{2}$ is an empirical parameter associated with non-uniform electric field in a CPW resonator, and $\delta_{\rm other}$ represents the photon-number-independent losses other than TLS loss (non-TLS loss) including residual resistance and radiation loss.
We assigned $F\delta_{\rm TLS, 0}$, $n_{\rm sat}$, $\beta_{2}$, and $\delta_{\rm other}$ as free fitting parameters.
The best-fit curves based on Eq.~(\ref{eq:1}) for Resonator 1 are representatively shown as solid curves in Fig.~\ref{fig:7}, and similar fitting results were obtained for the other resonators (see Appendix~\ref{app:D}).
The extracted $\delta_{\rm other}$ values for the four V-film structures are summarized in Fig.~\ref{fig:8}(b), where each value represents the average over all resonators on the corresponding chip.
The $\delta_{\rm other}$ value for the Si/Nb/V/Ta (Si/V/Ta) resonator, 1.94 $\times$ 10$^{-6}$ (0.76 $\times$ 10$^{-6}$), is significantly smaller than that for the Si/Nb/V (Si/V) resonator, 2.74 $\times$ 10$^{-6}$ (2.03 $\times$ 10$^{-6}$).
These results indicate that the Ta capping layer mitigates non-TLS losses in the V-based resonators.
With regard to the Nb buffer layer, comparison of the $\delta_{\rm other}$ values between the Si/Nb/V/Ta and Si/V/Ta (Si/Nb/V and Si/V) resonators suggests that V films grown on the Nb buffer layer contain more non-TLS loss sources in the bulk than those directly grown on the Si wafers.
 
\section{\label{sec:discussion} DISCUSSION}

By analyzing the observed  $Q_{\rm int}$ vs $\langle n_{\rm ph} \rangle$, we found that the Si/Nb/V/Ta (Si/V/Ta) resonators exhibited higher $Q_{\rm int}$, stronger $\langle n_{\rm ph} \rangle$ dependence of $Q_{\rm int}$, and lower $\delta_{\rm other}$ than the Si/Nb/V (Si/V) resonators.
These results suggest that non-TLS loss yielded at the V surface at the top of the V layers can be mitigated by the Ta capping layer, whereas surface oxidation of the Ta capping layer enhances TLS loss.
We assume that oxygen penetration into the bulk of the V layers is almost independent of the presence of the Ta capping layer.
Notably, even the $\delta_{\rm other}$ value for the Si/V/Ta resonator, 0.76 $\times$ 10$^{-6}$, which is the smallest among all resonators studied here, remains considerably larger than that reported for the $\alpha$-Ta resonators in Ref.~\onlinecite{Urade_Ta_APLMat}, in which $\delta_{\rm other}$ was obtained to be less than 0.045 $\times$ 10$^{-6}$. 
This indicates that substantial non-TLS loss sources may still reside in the bulk of the V layer and/or at the edge of the resonator conductors, where V surface is not covered with the Ta capping layer.
V oxidation, implied by the XPS and EDS analyses, is a prominent candidate for the origin of these non-TLS loss sources.
The XPS results shown in Fig.~\ref{fig:6} suggest that various V oxides, including V$_{2}$O$_{5}$, VO$_{2}$, and other suboxides, exist on the V surface.
Furthermore, as shown in Fig.~\ref{fig:5}, the EDS analysis reveal that O atoms are present not only at the surface but also in the bulk of the V layers.
As described in Sec.~\ref{sec:results-3}, oxygen may penetrate along grain boundaries, suggesting the formation of V oxides in the bulk region.
Among the V oxides, VO$_{2}$ and VO have been reported to exhibit semiconducting behavior and hopping transport of carriers at low temperatures \cite{Bharadwaja_APL, Singh_TSF}.
These characteristics could offer residual resistance, thereby contributing to $\delta_{\rm other}$ in the V-based resonators.
Another plausible candidate for the non-TLS loss sources is hydrogen incorporation in the V resonator structures.
Recent studies on Nb and $\alpha$-Ta resonators have suggested that hydrofluoric acid treatments during fabrication can cause hydrogen diffusion into metallic layers and the formation of Nb and Ta hydrides, which can give rise to power-independent loss \cite{Castanedo_Nb_AdvFM, Sung_Nb_PRMat, Lozano_Ta_AdvSci}.
In our V resonators, which were also fabricated using BHF treatment, the presence of V hydrides as a source of non-TLS loss is likewise plausible, considering the large hydrogen diffusion coefficient in V \cite{Suzuki_PRL, Luo_JPC} and the metallic conductivity of certain V hydrides \cite{Courtney1977, Li_InorgChem, Pan_IJHE}.

Finally, we discuss correlation between the structural properties of the V films and their microwave characteristics.
As summarized in Fig.~\ref{fig:8}, the resonators with the Si/Nb/V/Ta (Si/Nb/V) structure exhibited lower $Q_{\rm int}$ and higher $\delta_{\rm other}$ than those with the Si/V/Ta (Si/V) structure, whereas the XRD results shown in Fig.~\ref{fig:1} confirmed that the V films grown on the Nb buffer layer had higher lattice-orientation uniformity than those directly grown on the Si wafers. 
These results suggest that improved structural ordering of the V films dominantly mitigate loss, and that the presence of the Nb buffer layer introduces additional non-TLS loss sources in the film structures.
Possible additional loss sites associated with the Nb buffer layer include lattice defects at the Nb/V interface, disordered layers at the Si/Nb interface as can be seen in Figs.~\ref{fig:4}(d), and an increased density of grain boundaries.
In particular, we suspect that the increased grain-boundary density in the V layers is a dominant contributor to the enhanced non-TLS losses.
The AFM topographies shown in Figs.~\ref{fig:2}(a) and \ref{fig:2}(b) suggest that the V films grown on the Nb buffer layer contain smaller grains than those directly grown on the Si wafers.
The former is expected to have higher grain-boundary density, which may host V oxides, such as VO$_{2}$ and VO, and/or V hydrides, thereby increasing the density of non-TLS loss sources relative to the latter.
Further investigation of the oxidation and hydrogenation states of the surfaces, bulk, and grain boundaries is needed to obtain a detailed understanding of the microwave loss mechanism in superconducting V films.

\section{\label{sec:conclusion} CONCLUSION}

We fabricated and characterized 200-nm-thick V films sputtered on Si(001) wafers.
Nb and Ta were employed as materials for the 5-nm-thick buffer and capping layers, respectively, resulting in four multilayered structures: Nb/V/Ta, Nb/V, V/Ta, and V on the Si wafers.
XRD and AFM measurements revealed that the V films grown on the Nb buffer layer exhibited higher lattice-orientation uniformity and smaller grain size than those directly grown on the Si wafer.
$T_{\rm c}$ of all four V-film structures were confirmed to be close to 5.4 K, which is consistent with those of an epitaxial film and bulk of V \cite{Radebaugh_PR, Gutsche_TSF}. 
CPW resonators were fabricated from these V-film structures.
TEM observations confirmed the presence of the disordered layers at the Si/Nb and Si/V interfaces in the resonator structures, while EDS analysis suggested that oxygen atoms were distributed not only near the surface but also within the bulk of the V layers.
Furthermore, XPS analyses indicated that the V surface had various oxidation states, whereas Ta$_{2}$O$_{5}$ accounted for more than 70\% on the oxidation states of the Ta surfaces.

From the microwave measurements, we found that the presence of the Ta capping layer enhanced $Q_{\rm int}$, strengthened the $\langle n_{\rm ph} \rangle$ dependence of $Q_{\rm int}$, and reduced $\delta_{\rm other}$.
These results suggest that the Ta capping layer mitigates non-TLS losses yielded at the V surfaces, while the TLS loss associated with Ta oxides at the Ta surface may increase.
The non-TLS losses may be caused by residual resistance associated with V oxides exhibiting conductive channels \cite{Bharadwaja_APL, Singh_TSF}, existing in the surface and along grain boundaries in the bulk of the V films.
Although the Nb buffer layer was found to improve the crystallinity of the V films, it degraded $Q_{\rm int}$ and enhanced $\delta_{\rm other}$.
This indicates that improved lattice-orientation uniformity cannot dominantly mitigate loss in the V films, and that additional non-TLS-loss sources are introduced by the Nb buffer layer.
We speculate that the higher grain-boundary density in V films grown on Nb buffer layers leads to increased non-TLS loss due to V oxides and/or hydrides along the grain boundaries, when compared to films grown directly on Si wafers.
Although further investigation of the structural and chemical properties of surfaces, bulk regions, and grain boundaries is required to fully understand loss mechanisms in V films, the findings in this work provide important insights to bridge between material properties and loss mechanisms in superconducting films.

\begin{acknowledgments}
This paper was based on results obtained from a project, JPNP16007, commissioned by the New Energy and Industrial Technology Development Organization (NEDO), Japan. The devices/circuits/samples were fabricated in the Superconducting Quantum Circuit Fabrication Facility (Qufab) in National Institute of Advanced Industrial Science and Technology (AIST). A part of this work was supported by ``Advanced Research Infrastructure for Materials and Nanotechnology in Japan (ARIM)'' of the Ministry of Education, Culture, Sports, Science and Technology (MEXT). Proposal Number JPMXP1225AT0221. We acknowledge Dr. Toyofumi Ishikawa, Dr. Kosuke Mizuno, Dr. Akiyoshi Tomonaga, and Dr. Tomohiro Yamaji (AIST) for their support in the microwave measurement setup and Dr. Takayuki Nozaki and Dr. Kay Yakushiji (AIST) for their comments on film growth. We acknowledge Yamato Aoyama, Dr. Daisuke Ohori, and Prof. Kazuhiko Endo (Tohoku University) for useful discussion on the XPS analysis. Y.F. would like to thank Eiki Harashima, Michiyo Isaka, Katsumi Oono, Katsumasa Tashiro, and Tomokazu Yoshida (AIST) for instruction in the microfabrication.
\end{acknowledgments}

\appendix
\section{\label{app:A} FILM DEPOSITION AND CHARACTERIZATION}

To examine the growth of the V films, we prepared small pieces of Si wafers with the size of 20 mm $\times$ 10 mm, which were cut from a 4-inch, 300-$\mu$m-thick (001)-oriented Si wafer ($\rho \sim$ 1 k$\Omega$cm) supplied by Miyoshi Co., Ltd.
The wafers were firstly cleaned with 10:1 BHF (SA, Stella Chemifa) for 1 minute and rinsed with deionized water for 5 minutes.
Then, the wafers were loaded into a DC sputtering system (C-7100, Canon ANELVA) and the four V-film structures of Nb(5 nm)/V(200 nm)/Ta(5 nm), Nb(5 nm)/V(200 nm), V(200 nm)/Ta(5 nm), and V(200 nm) were separately deposited on individual wafers.
Each deposition process was performed without breaking the ultrahigh vacuum.
The parameters related to the deposition are summarized in Table~\ref{tab:sputtering}.
The crystallinity of the fabricated films was verified by XRD with Cu$K_{\alpha}$ radiation (45 kV, 200 mA), using an X-ray diffractometer (Smartlab, Rigaku) with a 2-dimensional detector (HyPix-3000, Rigaku).
The surface morphology of the fabricated films was evaluated by using an AFM system (NX10, Park Systems) and 1 $\mu$m $\times$ 1 $\mu$m areas in the surfaces were scanned with a non-contact mode.
We fabricated bar-shaped devices measuring 100 $\mu$m in length and 20 $\mu$m in width from the V-film structures, and measured $R$ under a DC current of 10 $\mu$A while sweeping $T$ from 3 to 300 K, using a physical property measurement system (PPMS, Quantum Design). 
The bar-shaped devices were fabricated near the corner on the resonator chips, and hence the microfabrication process for them was the same as that of the resonators described in Appendix~\ref{app:B}.
We also prepared a Si/Nb(5 nm) sample with a size of approximately 10 mm $\times$ 5 mm and performed a measurement of $R$ vs $T$ of it with the same setup as the measurements of the bar-shaped devices.
In this study, $R_{\rm N}$ is defined as the value of $R$ at the first measurement point where $dR/dT < 1$ ($\Omega$/K) after $dR/dT$ reaches its maximum.
$T_{\rm c}$ is defined as the value of $T$ at the first measurement point where $R$ exceeds 90 \% of $R_{\rm N}$.
\begin{table*}
\caption{\label{tab:sputtering} Sputtering-process parameters.}
\begin{ruledtabular}
\begin{tabular}{cccc}
Target&Nb&V&Ta\\
\hline
Specification&$\phi$110 mm, t4 mm, 4N&$\phi$110 mm, t4 mm, 3N&$\phi$164 mm, t10 mm, 5N \\
Base pressure (Pa)&3.0 $\times$ 10$^{-6}$& 3.0 $\times$ 10$^{-6}$& 1.9 $\times$ 10$^{-7}$ \\
Process gas&Ar, 30 sccm&Ar, 30 sccm&Kr, 15 sccm \\
Process pressure (Pa)&0.038&0.036&0.019 \\
DC power (W)&500&500&1000 \\
Stage rotation (rpm)&100&100&100 \\
Deposition rate (nm/s)&0.044&0.026&0.160 \\
\end{tabular}
\end{ruledtabular}
\end{table*}

\section{\label{app:B} MICROFABRICATION OF CPW RESONATORS}

We prepared double-side-polished 4-inch (001)-oriented Si wafers ($\rho \sim$ 15 k$\Omega$cm) supplied by Miyoshi Co., Ltd for the resonator fabrication.
The four different V-film structures were grown on the wafers through the same cleaning and deposition processes as those described in Appendix~\ref{app:A}.
The CPW resonators were formed by an i-line photolithography system (FPA-3030i5+, Canon) and an inductively-coupled-plasma (ICP) reactive ion etching system (RIE-400iPS, Samco) with a gas mixture of chlorine (Cl$_{2}$) and boron trichloride (BCl$_{3}$).
We fixed flow rate ratio of Cl$_{2}$ $:$ BCl$_{3}$, gas pressure, ICP radio frequency power, and bias power at 20 $:$ 15 (sccm), 0.7 Pa, 400 W, and 10 W, respectively, during the etching.
Then, the wafers were immersed in deionized water to remove residual chlorine compounds that can corrode the resonator structures, followed by removal of residual photoresist using O$_{2}$ plasma ashing and the stripping solution 104 (Tokyo Ohka Kogyo) with an ultrasonicator.
A photoresist was coated on the sample surface, and the samples were diced into 5 mm $\times$ 5 mm.
After dicing, the photoresist coated on the chip was removed by using acetone, isopropyl alcohol, and the stripping solution 104.
Each chip was cleaned by the BHF treatment described in Appendix~\ref{app:A} before being mounted in the dilution refrigerator system.

\section{\label{app:C} XPS MEASUREMENTS}
The XPS measurements were performed for the V, Ta, and Si surfaces on the resonator chips using an X-ray photoelectron spectrometer (KRATOS NOVA, SHIMADZU) with a monochromated Al $K$$_{\alpha}$ X-ray source (1486.6 eV) operated at 10 mA and 15 kV.
The peak fitting of the observed V 2$p$ (with O 1$s$) and Ta 4$f$ XPS spectra was performed using CasaXPS provided by Casa Software Ltd.
At each XPS measurement, the position of the C 1$s$ peak was identified for charge correction, and the binding energy was calibrated using the obtained C 1$s$ peak position.
In the peak fittings, we determined peak area, binding energy (peak position), and FWHM of V 2$p$ (and O 1$s$) and Ta 4$f$ peaks associated with each oxidation state as free parameters, while several parameters were constrained to reduce the number of free parameters.
The areas of V 2$p_{1/2}$ (Ta 4$f_{5/2}$) peaks were fixed at 0.5 (0.75) times those of V 2$p_{3/2}$ (Ta 4$f_{7/2}$), and various constraints are added into the peak positions and FWHM with reference to previous reports on the XPS analyses of the V and Ta surfaces \cite{Silversmit_JESRP, Brumbach_JVSTA}.

\begin{figure}
\includegraphics[width=8.5cm]{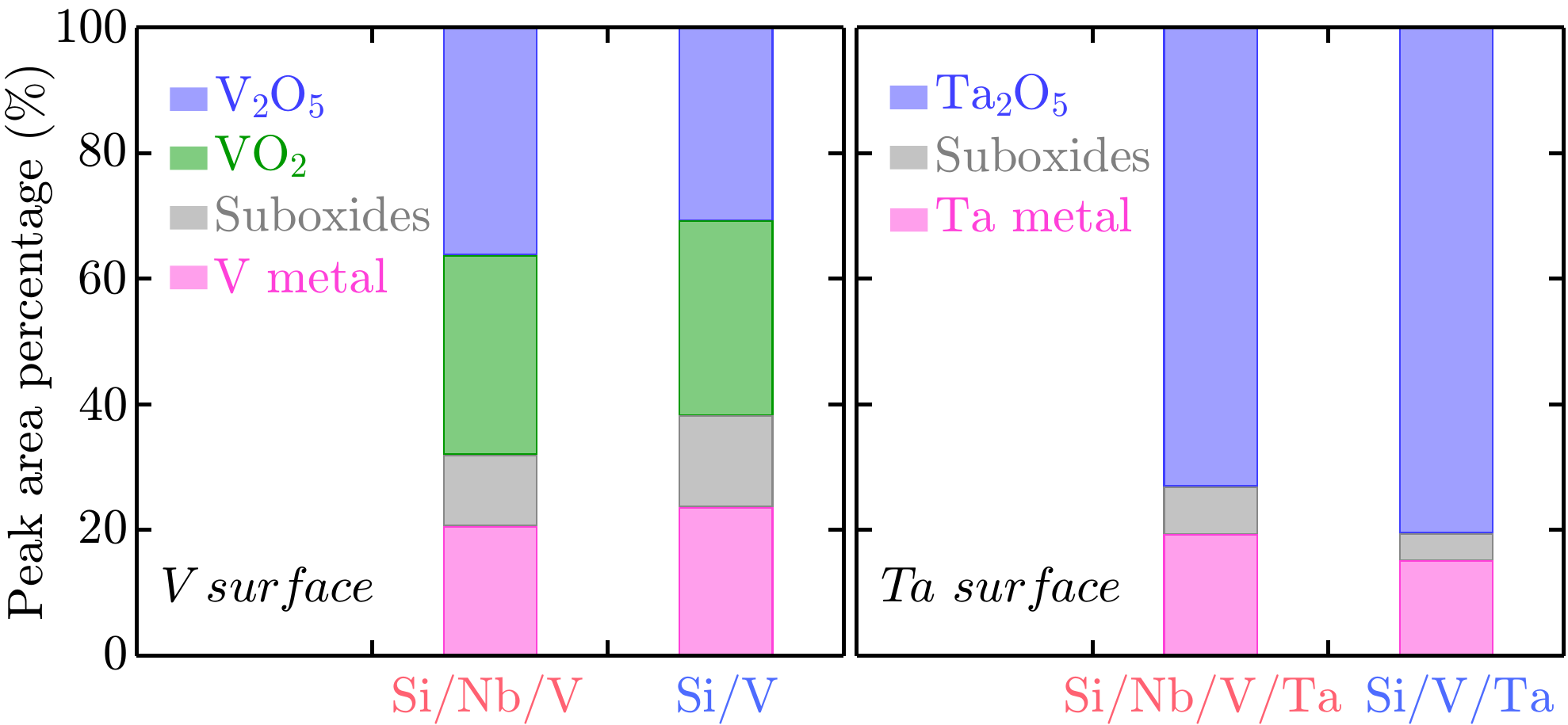}
\caption{\label{fig:9} Peak area percentage of surface oxides of V surfaces (left) and Ta surfaces (right) before the BHF treatment.}
\end{figure}

As mentioned in Sec.~\ref{sec:results-4}, we also performed the series of XPS measurements and analyses for the resonator chips before the BHF treatment.
Figure~\ref{fig:9} shows the peak area ratio of the oxidation-state components for the V and Ta surfaces.
As for the V surface, in comparison with the results after the BHF treatment shown in Fig.~\ref{fig:6}(e), it was suggested that the BHF treatment can decrease the contributions of the oxidation states of V$_{2}$O$_{5}$ and VO$_{2}$ and increase those of other suboxides and metallic V to the V surface.
In terms of surface oxide removal, the BHF treatment can be valid to remove V$_{2}$O$_{5}$ and VO$_{2}$ from the V surface, whereas its process conditions, such as immersion time and BHF concentration, should be modified to remove the other suboxides.
On the other hand, there was almost no difference in the oxidation states of the Ta surfaces before and after the BHF treatment, suggesting that the BHF treatment with the present process conditions is completely useless to remove the Ta oxides.

\section{\label{app:D} MICROWAVE MEASUREMENT SETUP AND ANALYSES}

\begin{table*}[b]
\caption{\label{tab:fc} Resonant frequency of all of the resonators.}
\begin{ruledtabular}
\begin{tabular}{ccccc}
&\multicolumn{4}{c}{$f_{\rm c}$ (GHz)}\\
Structure & Resonator 1 & Resonator 2 & Resonator 3 & Resonator 4\\ 
\hline
Si/Nb/V/Ta & 10.1125 & 10.4147 & 10.7369 & 11.0065\\
Si/Nb/V & 10.1281 & 10.4303 & 10.7531 & 11.0140\\
Si/V/Ta & 10.2576 & 10.5650 & 10.8909 & 11.1609\\
Si/V & 10.2094 & 10.5149 & 10.8384 & 11.1483\\
\end{tabular}
\end{ruledtabular}
\end{table*}

\begin{figure}
\includegraphics[width=8.5cm]{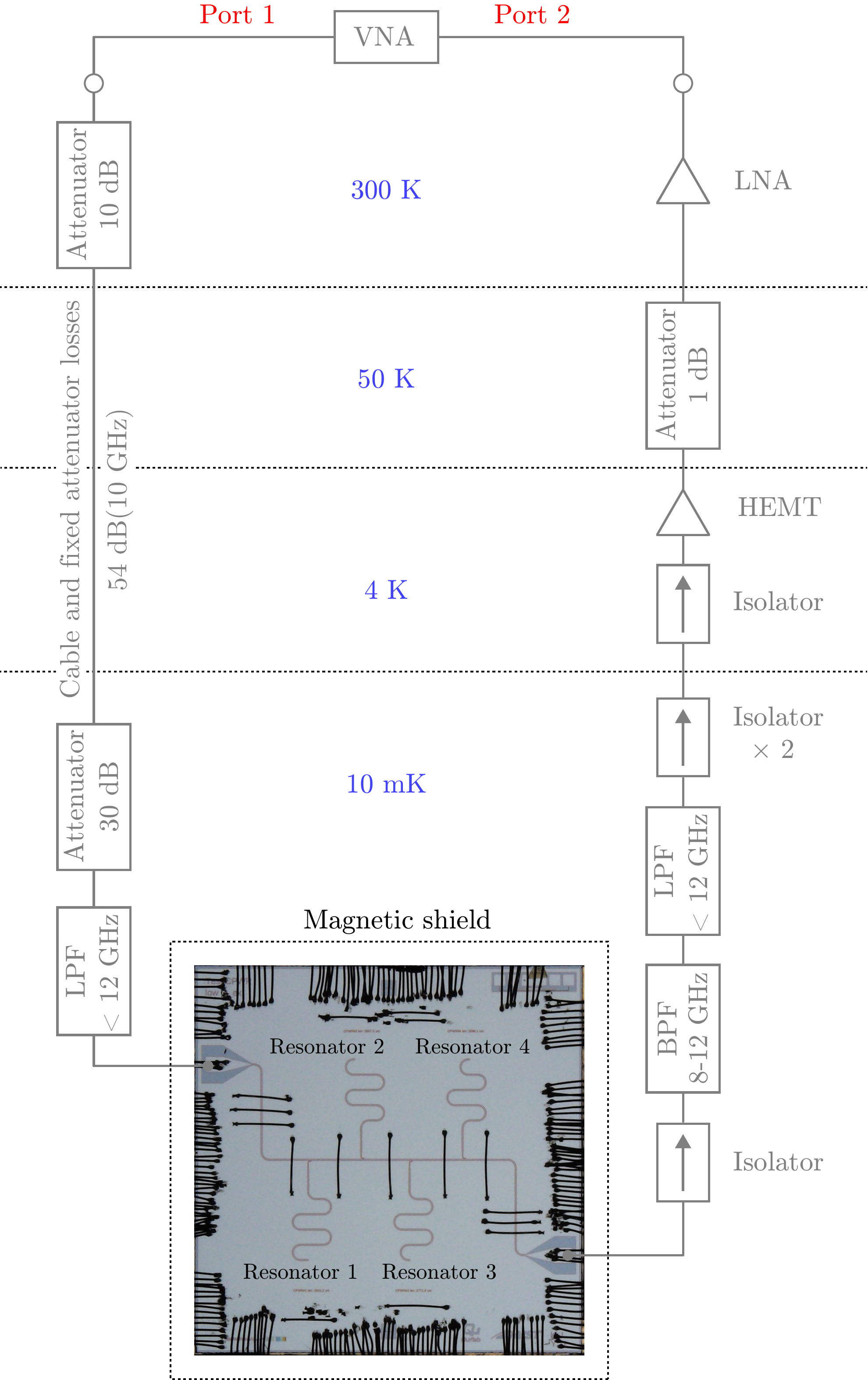}
\caption{\label{fig:10} Measurement setup with a representative optical micrograph of an Al-wire-bonded Si/V/Ta resonator chip. LNA, HEMT, LPF, and BPF denote a low-noise amplifier at RT, a high electron mobility transistor for a cryogenic amplifier, a low-pass and band-pass filters, respectively.}
\end{figure}
In Fig.~\ref{fig:10}, we show a representative micrograph of the Al-wire-bonded resonator chip and an overview of the measurement setup with a microwave circuit.
The Al-wire bonding was applied not only between the chip and the printed circuit board but also between the ground planes separated by the transmission line.
The sample packages embracing the resonator chip were enveloped by a magnetic shield and installed into a dilution refrigerator (LD250, Bluefors).
We used a VNA (M9805A, Keysight Technologies) placed on the room-temperature region to measure $S_{21}$.
The observed $S_{21}$ spectra were fitted with the following theoretical model to extract $Q_{\rm int}$.
\begin{equation}
\label{eq:2}
S_{21} = B\exp(j\phi_{\rm off}) \left\{1- \frac{{\frac{Q}{Q_{\rm ext}} - j\frac{2Qdf}{f_{\rm c}}}}{1 + j\frac{2Q(f - f_{\rm c})}{f_{\rm c}}} \right\},
\end{equation}
where $f$ is the probing frequency, $B$ is the background amplitude, $j$ is the imaginary unit, $\phi_{\rm off}$ is the phase offset, $Q$ is the loaded quality factor, $Q_{\rm ext}$ is the external (coupling) quality factor, and $df$ is the parameter in units of frequency characterizing asymmetry of shape of the $S_{21}$ spectrum.
From $Q$ and $Q_{\rm ext}$ obtained by fitting the observed $S_{21}$ response with Eq.~(\ref{eq:2}), $Q_{\rm int}$ can be derived using a following formula.
\begin{equation}
\label{eq:3}
Q_{\rm int} = \left(\frac{1}{Q} - \frac{1}{Q_{\rm ext}} \right)^{-1}.
\end{equation}
A representative fitting result of the $S_{21}$ response with Eq.~(\ref{eq:2}) was indicated by a solid curve in the inset of Fig.~\ref{fig:7}.
We confirmed $f_{\rm c}$ of each resonator in advance, before starting the $S_{21}$ measurement.
As summarized in Table~\ref{tab:fc}, the $f_{\rm c}$ values were confirmed to be 10 -- 11 GHz, which was as intended in designing.
We found that the resonator with the Nb buffer layer had lower $f_{\rm c}$ than that without the Nb buffer layer, implying that kinetic inductance of the former was larger than that of the latter.
The larger kinetic inductance may be due to the smaller grain size in the V film grown on the Nb buffer layer, compared with the film directly grown on the Si wafer.
\begin{figure*}
\includegraphics[width=17cm]{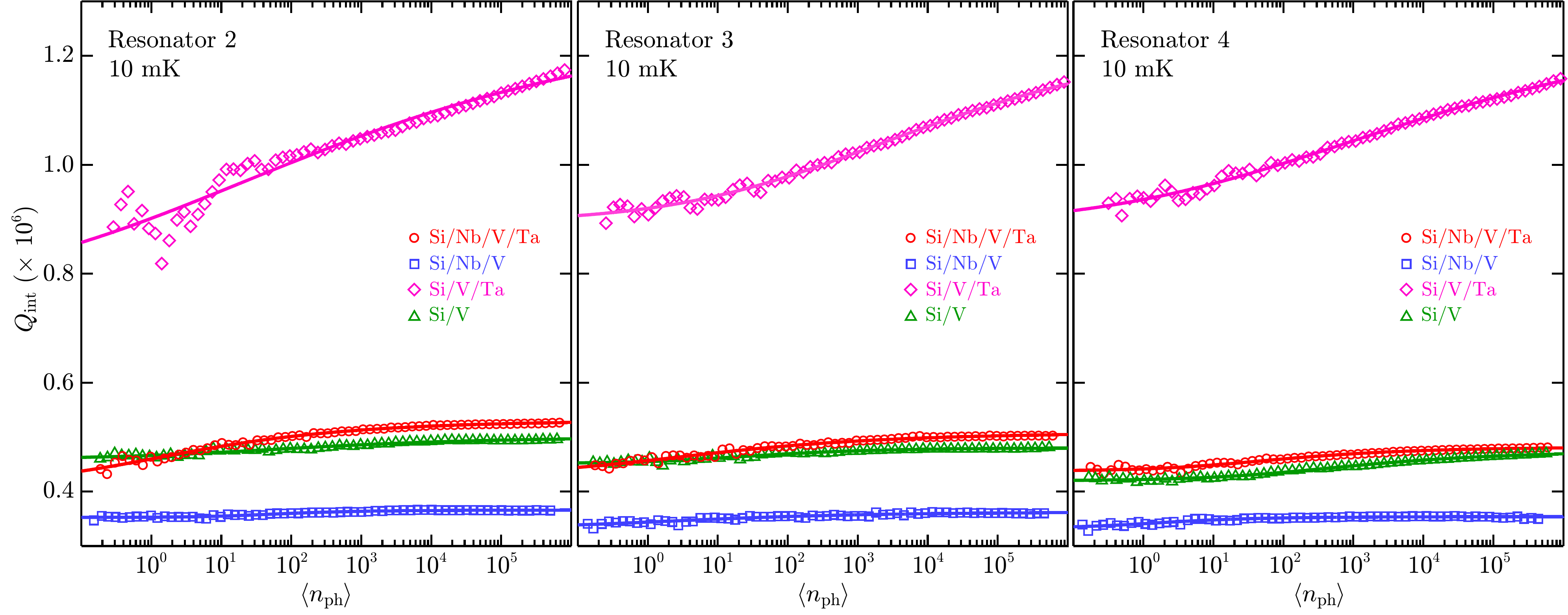}
\caption{\label{fig:11} $\langle n_{\rm ph} \rangle$ dependences of $Q_{\rm int}$ for Resonator 2, Resonator 3, and Resonator 4 with Si/Nb/V/Ta (red circle), Si/Nb/V (blue square), Si/V/Ta (magenta diamond), and Si/V (green triangle) structures. The data points and solid curves indicate the medians of repeated measurements and  the fitting results with Eq.~(\ref{eq:1}), respectively.}
\end{figure*}

\begin{table*}
\caption{\label{tab:fitting-Qint} Parameters obtained by fitting Eq.~(\ref{eq:1}) with the $\langle n_{\rm ph} \rangle$ dependences of $Q_{\rm int}$. The error indicates standard deviation.}
\begin{ruledtabular}
\begin{tabular}{cccccc}
Structure & Resonator & $\delta_{\rm other}$ (10$^{-6}$) & $F\delta_{\rm TLS,0}$ (10$^{-6}$) & $n_{\rm sat}$ & $\beta_{2}$\\
\hline
Si/Nb/V/Ta & Resonator 1 & 1.838 $\pm$ 0.009 & 0.337 $\pm$ 0.037 & 1.510 $\pm$ 1.117 & 0.460 $\pm$ 0.057\\
& Resonator 2 & 1.886 $\pm$ 0.008 & 0.537 $\pm$ 0.099 & 0.157 $\pm$ 0.186 & 0.486 $\pm$ 0.055\\
& Resonator 3 & 1.968 $\pm$ 0.009 & 0.355 $\pm$ 0.059 & 0.341 $\pm$ 0.404 & 0.438 $\pm$ 0.062\\
& Resonator 4 & 2.075 $\pm$ 0.006 & 0.209 $\pm$ 0.013 & 12.618 $\pm$ 4.261 & 0.596 $\pm$ 0.072\\
Si/Nb/V & Resonator 1 & 2.661 $\pm$ 0.006 & 0.126 $\pm$ 0.010 & 22.075 $\pm$ 8.667 & 1.062 $\pm$ 0.252\\
& Resonator 2 & 2.731 $\pm$ 0.003 & 0.104 $\pm$ 0.006 & 21.979 $\pm$ 6.270 & 0.933 $\pm$ 0.147\\
& Resonator 3 & 2.759 $\pm$ 0.013 & 0.242 $\pm$ 0.083 & 0.326 $\pm$ 0.784 & 0.451 $\pm$ 0.140\\
& Resonator 4 & 2.825 $\pm$ 0.004 & 0.173 $\pm$ 0.030 & 0.567 $\pm$ 0.465 & 0.858 $\pm$ 0.178\\
Si/V/Ta & Resonator 1 & 0.686 $\pm$ 0.036 & 0.439 $\pm$ 0.063 & 13.402 $\pm$ 9.763 & 0.207 $\pm$ 0.034\\
& Resonator 2 & 0.791 $\pm$ 0.062 & 0.531 $\pm$ 0.333 & 0.102 $\pm$ 0.668 & 0.252 $\pm$ 0.122\\
& Resonator 3 & 0.806 $\pm$ 0.016 & 0.314 $\pm$ 0.024 & 72.047 $\pm$ 19.510 & 0.326 $\pm$ 0.034\\
& Resonator 4 & 0.792 $\pm$ 0.021 & 0.338 $\pm$ 0.040 & 13.506 $\pm$ 8.770 & 0.515 $\pm$ 0.045\\
Si/V & Resonator 1 & 1.950 $\pm$ 0.013 & 0.412 $\pm$ 0.543 & 0.010 $\pm$ 0.084 & 0.421 $\pm$ 0.145\\
& Resonator 2 & 1.997 $\pm$ 0.011 & 0.177 $\pm$ 0.025 & 7.049 $\pm$ 5.703 & 0.408 $\pm$ 0.083\\
& Resonator 3 & 2.081 $\pm$ 0.004 & 0.132 $\pm$ 0.010 & 6.104 $\pm$ 2.571 & 0.649 $\pm$ 0.091\\
& Resonator 4 & 2.107 $\pm$ 0.010 & 0.274 $\pm$ 0.013 & 85.453 $\pm$ 15.491 & 0.515 $\pm$ 0.045\\
\end{tabular}
\end{ruledtabular}
\end{table*}

Figure~\ref{fig:11} shows $Q_{\rm int}$ as functions of $\langle n_{\rm ph} \rangle$ for Resonator 2-4 of the four V-film structures.
As with the Resonator 1 data shown in Fig.~\ref{fig:7}, the $Q_{\rm int}$ values shown in Fig.~\ref{fig:11} were the medians of the values obtained from repeated measurements at each data point.
The trends in magnitude relationships and $\langle n_{\rm ph} \rangle$ dependences of $Q_{\rm int}$ for Resonator 2-4 are clearly identical to those for Resonator 1 shown in Fig.~\ref{fig:7}.
By fitting Eq.~(\ref{eq:1}) with $\langle n_{\rm ph} \rangle$ dependences of $Q_{\rm int}$ as shown in Fig.~\ref{fig:7} and Fig.~\ref{fig:11}, we obtained the values of the fitting parameters, namely, $\delta_{\rm other}$, $F\delta_{\rm TLS,0}$, $n_{\rm sat}$, and $\beta_{2}$ for all resonators, Resonator 1-4, of the four V-film structures, as summarized in Table~\ref{tab:fitting-Qint}.
Focusing on $F\delta_{\rm TLS,0}$, the values averaged over Resonator 1-4 with the Si/Nb/V/Ta, Si/Nb/V, Si/V/Ta, and Si/V structures are obtained to be (0.360 $\pm$ 0.030) $\times$ 10$^{-6}$, (0.161 $\pm$ 0.022) $\times$ 10$^{-6}$, (0.406 $\pm$ 0.086) $\times$ 10$^{-6}$, and (0.249 $\pm$ 0.156) $\times$ 10$^{-6}$, respectively.
The resonator with the Si/Nb/V/Ta (Si/V/Ta) structure exhibits larger $F\delta_{\rm TLS,0}$ than that with the Si/Nb/V (Si/V) structure, suggesting that the Ta capping layer can enhance $F\delta_{\rm TLS,0}$ of the resonators.
This enhancement of $F\delta_{\rm TLS,0}$ might be originated from Ta oxides at the Ta surfaces, which can include more TLS sources than V oxides at the V surface.

\clearpage
\nocite{*}
\bibliography{manuscript}

@preamble{"\providecommand{\noopsort}[1]{}" #
   "\providecommand{\singleletter}[1]{#1}%"}

@article{Devoret_Science,
	author = {M. H. Devoret and R. J. Schoelkopf},
	journal = {Science},
	pages = {1169},
	title = {Superconducting circuits for quantum information: {A}n outlook},
	volume = {339},
	year = {2013}}

@article{Kjaergaard_Ann,
	author = {M. Kjaergaard and M. E. Schwartz and J. Braum{\"{u}}ller and P. Krantz and J. I.-J. Wang and S. Gustavsson and W. D. Oliver},
	journal = {Annu.\ Rev.\ Condens.\ Matter.\ Phys.},
	pages = {369},
	title = {Superconducting qubits: {C}urrent state of play},
	volume = {11},
	year = {2020}}

@article{Sage_JAP,
	author = {J. M. Sage and V. Bolkhovsky and W. D. Oliver and B. Turek and P. B. Welander},
	journal = {J.\ Appl.\ Phys.},
	pages = {063915},
	title = {Study of loss in superconducting coplanar waveguide resonators},
	volume = {109},
	year = {2011}}

@article{Oliver_MRS,
	author = {W. D. Oliver and P. B. Welander},
	journal = {MRS Bull.},
	pages = {816},
	title = {Materials in superconducting quantum bits},
	volume = {38},
	year = {2013}}

@article{McRae_Rev,
	author = {C. R. H. McRae and H. Wang and J. Gao and M. R. Vissers and T. Brecht and A. Dunsworth and D. P. Pappas and J. Mutus},
	journal = {Rev.\ Sci.\ Instrum.},
	pages = {091101},
	title = {Materials loss measurements using superconducting microwave resonators},
	volume = {91},
	year = {2020}}

@article{Leon_Science,
	author = {N. P. de Leon and K. M. Itoh and D. Kim and K. K. Mehta and T. E. Northup and H. Paik and B. S. Palmer and N. Samarth and S. Sangtawesin and D. W. Steuerman},
	journal = {Science},
	pages = {253},
	title = {Materials challenges and opportunities for quantum computing hardware},
	volume = {372},
	year = {2021}}

@article{Siddiqi_NatRev,
	author = {I. Siddiqi},
	journal = {Nat.\ Rev.\ Mater.},
	pages = {875},
	title = {Engineering high-coherence superconducting qubits},
	volume = {6},
	year = {2021}}

@article{Connell_Al_APL,
	author = {A. D. O'Connell and M. Ansmann and R. C. Bialczak and M. Hofheinz and N. Katz and E. Lucero and C. McKenney and M. Neeley and H. Wang and E. M. Weig and A. N. Cleland and J. M. Martinis},
	journal = {Appl.\ Phys.\ Lett.},
	pages = {112903},
	title = {Microwave dielectric loss at single photon energies and millikelvin temperatures},
	volume = {92},
	year = {2008}}

@article{Wang_Al_APL,
	author = {H. Wang and M. Hofheinz and J. Wenner and M. Ansmann and R. C. Bialczak and M. Lenander and E. Lucero and M. Neeley and A. D. O'Connell and D. Sank and M. Weides and A. N. Cleland and J. M. Martinis},
	journal = {Appl.\ Phys.\ Lett.},
	pages = {233508},
	title = {Improving the coherence time of superconducting coplanar resonators},
	volume = {95},
	year = {2009}}

@article{Megrant_Al_APL,
	author = {A. Megrant and C. Neill and R. Barends and B. Chiaro and Yu Chen and L. Feigl and J. Kelly and E. Lucero and M. Mariantoni and P. J. J. O'Malley and D. Sank and A. Vainsencher and J. Wenner and T. C. White and Y. Yin and J. Zhao and C. J. Palmstr{\o}m and J. M. Martinis and A. N. Cleland},
	journal = {Appl.\ Phys.\ Lett.},
	pages = {113510},
	title = {Planar superconducting resonators with internal quality factors above one million},
	volume = {100},
	year = {2012}}

@article{Richardson_Al_SST,
	author = {C. J. K. Richardson and N. P. Siwak and J. Hackley and Z. K. Keane and J. E. Robinson and B. Arey and I. Arslan and B. S. Palmer},
	journal = {Supercond.\ Sci.\ Technol.},
	pages = {064003},
	title = {Fabrication artifacts and parallel loss channels in metamorphic epitaxial aluminum superconducting resonators},
	volume = {29},
	year = {2016}}

@article{Dunsworth_Al_APL,
	author = {A. Dunsworth and A. Megrant and C. Quintana and Z. Chen and R. Barends and B. Burkett and B. Foxen and Y. Chen and B. Chiaro and A. Fowler and R. Graff and E. Jeffrey and J. Kelly and E. Lucero and J. Y. Mutus and M. Neeley and C. Neill and P. Roushan and D. Sank and A. Vainsencher and others},
	journal = {Appl.\ Phys.\ Lett.},
	pages = {022601},
	title = {Characterization and reduction of capacitive loss induced by sub-micron {J}osephson junction fabrication in superconducting qubits},
	volume = {111},
	year = {2017}}

@article{Grunhaupt_Al_PRL,
	author = {L. Gr{\"{u}}nhaupt and N. Maleeva and S. T. Skacel and M. Calvo and F. Levy-Bertrand and A. V. Ustinov and H. Rotzinger and A. Monfardini and G. Catelani and I. M. Pop},
	journal = {Phys.\ Rev.\ Lett.},
	pages = {117001},
	title = {Loss mechanisms and quasiparticle dynamics in superconducting microwave resonators made of thin-film granular aluminum},
	volume = {121},
	year = {2018}}

@article{Earnest_Al_SST,
	author = {C. T. Earnest and J. H. B{\'{e}}janin and T. G. McConkey and E. A. Peters and A. Korinek and H. Yuan and M. Mariantoni},
	journal = {Supercond.\ Sci.\ Technol.},
	pages = {125013},
	title = {Substrate surface engineering for high-quality silicon/aluminum superconducting resonators},
	volume = {31},
	year = {2018}}

@article{Un_Al_SciAdv,
	author = {S. Un and S. de Graaf and P. Bertet and S. Kubatkin and A. Danilov},
	journal = {Sci.\ Adv.},
	pages = {eabm6169},
	title = {On the nature of decoherence in quantum circuits: Revealing the structural motif of the surface radicals in  $\alpha$-{A}l$_{2}${O}$_{3}$},
	volume = {8},
	year = {2022}}

@article{Kumar_Nb_APL,
	author = {S. Kumar and J. Gao and J. Zmuidzinas and B. A. Mazin and H. G. LeDuc and P. K. Day},
	journal = {Appl.\ Phys.\ Lett.},
	pages = {123503},
	title = {Temperature dependence of the frequency and noise of superconducting coplanar waveguide resonators},
	volume = {92},
	year = {2008}}

@article{Macha_Nb_APL,
	author = {P. Macha and S. H. W. van der Ploeg and G. Oelsner and E. Il'ichev and H.-G. Meyer and S. W{\"{u}}nsch and M. Siegel},
	journal = {Appl.\ Phys.\ Lett.},
	pages = {062503},
	title = {Losses in coplanar waveguide resonators at millikelvin temperatures},
	volume = {96},
	year = {2010}}

@article{Burnett_Nb_SST,
	author = {J. Burnett and L. Faoro and T. Lindstr{\"{o}}m},
	journal = {Supercond.\ Sci.\ Technol.},
	pages = {044008},
	title = {Analysis of high quality superconducting resonators: consequences for {T}{L}{S} properties in amorphous oxides},
	volume = {29},
	year = {2016}}

@article{Gambetta_Nb_IEEE,
	author = {J. M. Gambetta and C. E. Murray and Y.-K.-K. Fung and D. T. McClure and O. Dial and W. Shanks and J. W. Sleight and M. Steffen},
	journal = {IEEE \ Trans.\ Appl.\ Supercond.},
	pages = {1700205},
	title = {Investigating surface loss effects in superconducting transmon qubits},
	volume = {27},
	year = {2016}}

@article{Verjauw_Nb_PRAP,
	author = {J. Verjauw and A. Poto{\v{c}}nik and M. Mongillo and R. Acharya and F. Mohiyaddin and G. Simion and A. Pacco and Ts. Ivanov and D. Wan and A. Vanleenhove and L. Souriau and J. Jussot and A. Thiam and J. Swerts and X. Piao and S. Couet and M. Heyns and B. Govoreanu and I. Radu},
	journal = {Phys.\ Rev.\ Applied},
	pages = {014018},
	title = {Investigation of microwave loss induced by oxide regrowth in high-${Q}$ niobium resonators},
	volume = {16},
	year = {2021}}

@article{Altoe_Nb_PRXQ,
	author = {M. V. P. Alto{\'{e}} and A. Banerjee and C. Berk and A. Hajr and A. Schwartzberg and C. Song and M. Alghadeer and S. Aloni and M. J. Elowson and J. M. Kreikebaum and E. K. Wong and S. M. Griﬃn and S. Rao and A. Weber-Bargioni and A. M. Minor and D. I. Santiago and S. Cabrini and I. Siddiqi and D. F. Ogletree},
	journal = {PRX Quantum},
	pages = {020312},
	title = {Localization and mitigation of loss in niobium superconducting circuits},
	volume = {3},
	year = {2022}}

@article{Castanedo_Nb_AdvFM,
	author = {C. G. Torres-Castanedo and D. P. Goronzy and T. Pham and A. McFadden and N. Materise and P. M. Das and M. Cheng and D. Lebedev and S. M. Ribet and M. J. Walker and others},
	journal = {Adv.\ Funct.\ Mater.},
	pages = {2401365},
	title = {Formation and microwave losses of hydrides in superconducting niobium thin films resulting from fluoride chemical processing},
	volume = {34},
	year = {2024}}

@article{Sung_Nb_PRMat,
	author = {Z. Sung and D. Bafia and A. Cano and A. Murthy and J. Lee and M. J. Reagor and J. Rubio-Zuazo and A. Grassellino and A. Romanenko},
	journal = {Phys.\ Rev.\ Materials},
	pages = {016201},
	title = {Formation of niobium hydride precipitates in superconducting qubits},
	volume = {10},
	year = {2026}}

@article{Place_Ta_NatCom,
	author = {A. P. M. Place and L. V. H. Rodgers and P. Mundada and B. M. Smitham and M. Fitzpatrick and Z. Leng and A. Premkumar and J. Bryon and A. Vrajitoarea and S. Sussman and others},
	journal = {Nat.\ Commun.},
	pages = {1779},
	title = {New material platform for superconducting transmon qubits with coherence times exceeding 0.3 milliseconds},
	volume = {12},
	year = {2021}}

@article{Wang_Ta_npjQ,
	author = {C. Wang and X. Li and H. Xu and Z. Li and J. Wang and Z. Yang and Z. Mi and X. Liang and T. Su and C. Yang and others},
	journal = {npj\ Quantum\ Inf.},
	pages = {3},
	title = {Towards practical quantum computers: {T}ransmon qubit with a lifetime approaching 0.5 milliseconds},
	volume = {8},
	year = {2022}}

@article{Crowley_Ta_PRX,
	author = {K. D. Crowley and R. A. McLellan and A. Dutta and N. Shumiya and A. P. M. Place and X. H. Le and Y. Gang and T. Madhavan and M. P. Bland and R. Chang and others},
	journal = {Phys.\ Rev.\ X},
	pages = {041005},
	title = {Disentangling losses in tantalum superconducting circuits},
	volume = {13},
	year = {2023}}

@article{Urade_Ta_APLMat,
	author = {Y. Urade and K. Yakushiji and M. Tsujimoto and T. Yamada and K. Makise and W. Mizubayashi and K. Inomata},
	journal = {APL\ Mater.},
	pages = {021132},
	title = {Microwave characterization of tantalum superconducting resonators on silicon substrate with niobium buffer layer},
	volume = {12},
	year = {2024}}

@article{Lozano_Ta_MatQT,
	author = {D. P. Lozano and M. Mongillo and X. Piao and S. Couet and D. Wan and Y. Canvel and A. M. Vadiraj and T. Ivanov and J. Verjauw and R. Acharya and others},
	journal = {Mater.\ Quantum\ Technol.},
	pages = {025801},
	title = {Low-loss $\alpha$-tantalum coplanar waveguide resonators on silicon wafers: {F}abrication, characterization and surface modification},
	volume = {4},
	year = {2024}}

@article{Lozano_Ta_AdvSci,
	author = {D. P. Lozano and M. Mongillo and B. Raes and Y. Canvel and S. Massar and A. M. Vadiraj and T. Ivanov and R. Acharya and J. V. Damme and J. V. de Vondel and others},
	journal = {Adv.\ Sci.},
	pages = {e09244},
	title = {Reversing Hydrogen-Related Loss in $\alpha$-{T}a Thin Films for Quantum Device Fabrication},
	volume = {12},
	year = {2025}}

@article{Vissers_TiN_APL,
	author = {M. R. Vissers and J. Gao and D. S. Wisbey and D. A. Hite and C. C. Tsuei and A. D. Corcoles and M. Steffen and D. P. Pappas},
	journal = {Appl.\ Phys.\ Lett.},
	pages = {232509},
	title = {Low loss superconducting titanium nitride coplanar waveguide resonators},
	volume = {97},
	year = {2010}}

@article{Ohya_TiN_SST,
	author = {S. Ohya and B. Chiaro and A. Megrant and C. Neill and R. Barends and Y. Chen and J. Kelly and D. Low and J. Mutus and P. J. J. O'Malley and others},
	journal = {Supercond.\ Sci.\ Technol.},
	pages = {082602},
	title = {Room temperature deposition of sputtered {T}i{N} films for superconducting coplanar waveguide resonators},
	volume = {27},
	year = {2013}}

@article{Bruno_NbTiN_APL,
	author = {A. Bruno and G. de Lange and S. Asaad and K. L. van der Enden and N. K. Langford and L. DiCarlo},
	journal = {Appl.\ Phys.\ Lett.},
	pages = {182601},
	title = {Reducing intrinsic loss in superconducting resonators by surface treatment and deep etching of silicon substrates},
	volume = {106},
	year = {2015}}

@article{deGraaf_NbN_NatCom,
	author = {S. E. de Graaf and L. Faoro and J. Burnett and A. A. Adamyan and A. Y. Tzalenchuk and S. E. Kubatkin and T. Lindstr{\"{o}}m and A. V. Danilov},
	journal = {Nat.\ Commun.},
	pages = {1143},
	title = {Suppression of low-frequency charge noise in superconducting resonators by surface spin desorption},
	volume = {9},
	year = {2018}}

@article{Terai_NbN_ComMat,
	author = {S. Kim and H. Terai and T. Yamashita and W. Qiu and T. Fuse and F. Yoshihara and S. Ashhab and K. Inomata and K. Semba},
	journal = {Commun.\ Mater.},
	pages = {98},
	title = {Enhanced coherence of all-nitride superconducting qubits epitaxially grown on silicon substrate},
	volume = {2},
	year = {2021}}

@article{Premkumar_Nb_ComMat,
	author = {A. Premkumar and C. Weiland and S. Hwang and B. J{\"{a}}ck and A. P. M. Place and I. Waluyo and A. Hunt and V. Bisogni and J. Pelliciari and A. Barbour and M. S. Miller and P. Russo and F. Camino and K. Kisslinger and X. Tong and M. S. Hybertsen and A. A. Houck and I. Jarrige},
	journal = {Commun.\ Mater.},
	pages = {72},
	title = {Microscopic relaxation channels in materials for superconducting qubits},
	volume = {2},
	year = {2021}}

@article{Alegria_Ta_APL,
	author = {L. D. Alegria and D. M. Tennant and K. R. Chaves and J. R. I. Lee and S. R. O'Kelley and Y. J. Rosen and J. L. DuBois},
	journal = {Appl.\ Phys.\ Lett.},
	pages = {062601},
	title = {Two-level systems in nucleated and non-nucleated epitaxial alpha-tantalum films},
	volume = {123},
	year = {2024}}

@article{Jones_Ta_JAP,
	author = {S. G. Jones and N. Materise and K. W. Leung and J. C. Weber and B. D. Isakov and X. Chen and J. Zheng and A. Gyenis and B. Jaeck and C. R. H. McRae},
	journal = {J.\ Appl.\ Phys.},
	pages = {144402},
	title = {Grain size in low loss superconducting {T}a thin films on c axis sapphire},
	volume = {134},
	year = {2023}}

@article{Bland_Ta_Nature,
	author = {M. P. Bland and F. Bahrami and J. G. C. Martinez and P. H. Prestegaard and B. M. Smitham and A. Joshi and E. Hedrick and S. Kumar and A. Yang and A. C. Pakpour-Tabrizi and A. Jindal and R. D. Chang and G. Cheng and N. Yao and R. J. Cava and N. P. de Leon and A. A. Houck},
	journal = {Nature},
	pages = {343},
	title = {Millisecond lifetimes and coherence times in 2{D} transmon qubits},
	volume = {647},
	year = {2025}}

@article{Gao_TiN_PRMat,
	author = {R. Gao and W. Yu and H. Deng and H.-S. Ku and Z. Li and M. Wang and X. Miao and Y. Lin and C. Deng},
	journal = {Phys.\ Rev.\ Materials},
	pages = {036202},
	title = {Epitaxial titanium nitride microwave resonators: structural, chemical, electrical, and microwave properties},
	volume = {6},
	year = {2022}}

@conference{Singer_Ta_QCE,
	author = {M. Singer and B. Schoof and H. Gupta and D. Zahn and J. Weber and M. Tornow},
	booktitle = {2024 IEEE International Conference on Quantum Computing and Engineering (QCE), Montreal, QC, Canada},
	pages = {1197},
	title = {Tantalum thin films sputtered on silicon and on different seed layers: {M}aterial characterization and coplanar waveguide resonator performance},
	year = {2024}}

@article{Schijndel_Ta_PRAP,
	author = {T. A. J. van Schijndel and A. P. McFadden and A. N. Engel and J. T. Dong and W. J. Y.-Parre{\~{n}}o and M. Parthasarathy and R. W. Simmonds and C. J. Palmstr{\o}m},
	journal = {Phys.\ Rev.\ Applied},
	pages = {034025},
	title = {Cryogenic growth of tantalum thin films for low-loss superconducting circuits},
	volume = {23},
	year = {2025}}

@article{Marcaud_Ta_ComMat,
	author = {G. Marcaud and D. Perello and C. Chen and E. Umbarkar and C. Weiland and J. Gao and S. Diez and V. Ly and N. Mahuli and N. D'Souza and Y. He and S. Aghaeimeibodi and R. Resnick and C. Jaye and A. K. Rumaiz and D. A. Fischer and M. Hunt and O. Painter and I. Jarrige},
	journal = {Commun.\ Mater.},
	pages = {182},
	title = {Low-loss superconducting resonators fabricated from tantalum films grown at room temperature},
	volume = {6},
	year = {2025}}

@article{Dhundhwal_Ta_APL,
	author = {R. Dhundhwal and H. Duan and L. Brauch and S. Arabi and D. Fuchs and A.-A. Haghighirad and A. Welle and F. Scharwaechter and S. Pal and M. Scheffler and J. Palomo and Z. Leghtas and A. Murani and H. Hahn and J. Aghassi-Hagmann and C. K{\"{u}}bel and W. Wulfhekel and I. M. Pop and T. Reisinger},
	journal = {Appl.\ Phys.\ Lett.},
	pages = {214005},
	title = {High-quality superconducting tantalum resonators with beta phase defects},
	volume = {127},
	year = {2025}}

@article{Ando_JMSJ,
	author = {F. Ando and D. Kan and Y. Shiota and T. Moriyama and Y. Shimakawa and T. Ono},
	journal = {J.\ Magn.\ Soc.\ Jpn.},
	pages = {17},
	title = {Fabrication of noncentrosymmetric {N}b/{V}/{T}a superlattice and its superconductivity},
	volume = {43},
	year = {2019}}

@article{Ando_Nature,
	author = {F. Ando and Y. Miyasaka and T. Li and J. Ishizuka and T. Arakawa and Y. Shiota and T. Moriyama and Y. Yanase and T. Ono},
	journal = {Nature},
	pages = {373},
	title = {Observation of superconducting diode effect},
	volume = {584},
	year = {2020}}

@article{Radebaugh_PR,
	author = {R. Radebaugh and P. H. Keesom},
	journal = {Phys.\ Rev.},
	pages = {149},
	title = {Low-temperature thermodynamic properties of vanadium. {I}. superconducting and normal states},
	volume = {209},
	year = {1966}}

@article{Gutsche_TSF,
	author = {M. Gutsche and H. Kraus and J. Jochum and B. Kemmather and G. Gutekunst},
	journal = {Thin Solid Films},
	pages = {18},
	title = {Growth and characterization of epitaxial vanadium films},
	volume = {248},
	year = {1994}}

@article{Werthamer_PR,
	author = {N. R. Werthamer},
	journal = {Phys.\ Rev.},
	pages = {2440},
	title = {Theory of the superconducting transition temperature and energy gap function of superposed metal films},
	volume = {132},
	year = {1963}}

@article{Gennes_RMP,
	author = {P. G. de Gennes},
	journal = {Rev.\ Mod.\ Phys.},
	pages = {225},
	title = {Boundary effects in superconductors},
	volume = {36},
	year = {1964}}

@article{Arrabal_JLTP,
	author = {R. Gonzalez-Arrabal and A. Cam{\'{o}}n and M. Parra-Border{\'{i}}as and L. Fabrega and J. Anguita and J. Ses{\'{e}} and F. Briones},
	journal = {J.\ Low.\ Temp.\ Phys.},
	pages = {239},
	title = {{M}o/{A}u bilayers deposited by sputtering at room temperature for transition edge sensors fabrication},
	volume = {151},
	year = {2008}}

@article{Weber_SST,
	author = {J. C. Weber and K. M. Morgan and D. Yan and C. G. Pappas and A. L. Wessels and G. C. O'Neil and D. A. Bennett and G. C. Hilton and D. S. Swetz and J. N. Ullom and D. R. Schmidt},
	journal = {Supercond.\ Sci.\ Technol.},
	pages = {115002},
	title = {Development of a transition-edge sensor bilayer process providing new modalities for critical temperature control},
	volume = {33},
	year = {2020}}

@article{Finnemore_PR,
	author = {D. K. Finnemore and T. F. Stromberg and C. A. Swenson},
	journal = {Phys.\ Rev.},
	pages = {231},
	title = {Superconducting Properties of High-Purity Niobium},
	volume = {149},
	year = {1966}}

@article{Mayadas_JAP,
	author = {A. F. Mayadas and R. B. Laibowitz and J. J. Cuomo},
	journal = {J.\ Appl.\ Phys.},
	pages = {1287},
	title = {Electrical characteristics of rf‐sputtered single‐crystal niobium films},
	volume = {43},
	year = {1972}}

@article{Wolf_JVST,
	author = {S. A. Wolf and J. J. Kennedy and M. Nisenoff},
	journal = {J.\ Vac.\ Sci.\ Technol.},
	pages = {145},
	title = {Properties of superconducting rf sputtered ultrathin films of Nb},
	volume = {13},
	year = {1976}}

@article{Kodama_JAP,
	author = {J. Kodama and M. Itoh and H. Hirai},
	journal = {J.\ Appl.\ Phys.},
	pages = {4050},
	title = {Superconducting transition temperature versus thickness of {N}b film on various substrates},
	volume = {54},
	year = {1983}}

@article{Minhaj_PRB,
	author = {M. S. M. Minhaj and S. Meepagala and J. T. Chen and L. E. Wenger},
	journal = {Phys.\ Rev.\ B},
	pages = {15235},
	title = {Thickness dependence on the superconducting properties of thin {N}b films},
	volume = {49},
	year = {1994}}

@article{Read_APL,
	author = {M. H. Read and C. Altman},
	journal = {Appl.\ Phys.\ Lett.},
	pages = {51},
	title = {A NEW STRUCTURE IN TANTALUM THIN FILMS},
	volume = {7},
	year = {1965}}

@article{Silversmit_JESRP,
	author = {G. Silversmit and D. Depla and H. Poelman and G. B. Marin and R. {De Gryse}},
	journal = {J.\ Electron Spectrosc.\ Relat.\ Phenom.},
	pages = {167},
	title = {Determination of the {V}2$p$ {X}{P}{S} binding energies for different vanadium oxidation states ({V}$^{5+}$ to {V}$^{0+}$)},
	volume = {135},
	year = {2004}}

@article{Biesinger_ASS,
	author = {M. C. Biesinger and L. W. M. Lau and A. R. Gerson and R. St.C. Smart},
	journal = {Appl.\ Surf.\ Sci.},
	pages = {887},
	title = {Resolving surface chemical states in {X}{P}{S} analysis of first row transition metals, oxides and hydroxides: {S}c, {T}i, {V}, {C}u and {Z}n},
	volume = {257},
	year = {2010}}

@article{Brumbach_JVSTA,
	author = {M. T. Brumbach and P. R. Mickel and A. J. Lohn and A. J. Mirabal and M. A. Kalan and J. E. Stevens and M. J. Marinella},
	journal = {J.\ Vac.\ Sci.\ Technol.\ A},
	pages = {051403},
	title = {Evaluating tantalum oxide stoichiometry and oxidation states for optimal memristor performance},
	volume = {32},
	year = {2014}}

@article{McLellan_AdvSci,
	author = {R. A. McLellan and A. Dutta and C. Zhou and Y. Jia and C. Weiland and X. Gui and A. P. M. Place and K. D. Crowley and X. H. Le and T. Madhavan and others},
	journal = {Adv.\ Sci.},
	pages = {2300921},
	title = {Chemical profiles of the oxides on tantalum in state of the art superconducting circuits},
	volume = {10},
	year = {2023}}

@article{Bruno_APL,
	author = {A. Bruno and G. de Lange and S. Asaad and K. L. van der Enden and N. K. Langford and L. DiCarlo},
	journal = {Appl.\ Phys.\ Lett.},
	pages = {182601},
	title = {Reducing intrinsic loss in superconducting resonators by surface treatment and deep etching of silicon substrates},
	volume = {106},
	year = {2015}}

@article{Khalil_JAP,
	author = {M. S. Khalil and M. J. A. Stoutimore and F. C. Wellstood and K. D. Osborn},
	journal = {J.\ Appl.\ Phys.},
	pages = {054510},
	title = {An analysis method for asymmetric resonator transmission applied to superconducting devices},
	volume = {111},
	year = {2012}}

@article{Geerlings_APL,
	author = {K. Geerlings and S. Shankar and E. Edwards and L. Frunzio and R. J. Schoelkopf and M. H. Devoret},
	journal = {Appl.\ Phys.\ Lett.},
	pages = {192601},
	title = {Improving the quality factor of microwave compact resonators by optimizing their geometrical parameters},
	volume = {100},
	year = {2012}}

@article{Wang_APL,
	author = {H. Wang and M. Hofheinz and J. Wenner and M. Ansmann and R. C. Bialczak and M. Lenander and E. Lucero and M. Neeley and A. D. O'Connell and D. Sank and M. Weides and A. N. Cleland and J. M. Martinis},
	journal = {Appl.\ Phys.\ Lett.},
	pages = {233508},
	title = {Improving the coherence time of superconducting coplanar resonators},
	volume = {95},
	year = {2009}}

@article{Bharadwaja_APL,
	author = {S. S. N. Bharadwaja and C. Venkatasubramanian and N. Fieldhouse and S. Ashok and M. W. Horn and T. N. Jackson},
	journal = {Appl.\ Phys.\ Lett.},
	pages = {222110},
	title = {Low temperature charge carrier hopping transport mechanism in vanadium oxide thin films grown using pulsed dc sputtering},
	volume = {94},
	year = {2009}}

@article{Singh_TSF,
	author = {A. K. Singh and S. Kumari and H. K. Singh and P. K. Siwach},
	journal = {Thin Solid Films},
	pages = {140773},
	title = {Substrate-induced modulation of metal-insulator transition and low-temperature charge transport in radio frequency sputtered {V}{O}$_{2}$ films},
	volume = {826},
	year = {2025}}

@article{Suzuki_PRL,
	author = {T. Suzuki and H. Namazue and S. Koike and H. Hayakawa},
	journal = {Phys.\ Rev.\ Lett.},
	pages = {798},
	title = {Superdiffusion of 4{$T$}-Hydrogen in Vanadium},
	volume = {51},
	year = {1983}}

@article{Luo_JPC,
	author = {J. Luo and H.-B. Zhou and Y.-L. Liu and L.-J. Gui and S. Jin and Y. Zhang and G.-H. Lu},
	journal = {J.\ Phys.:\ Condens.\ Matter},
	pages = {135501},
	title = {Dissolution, diffusion and permeation behavior of hydrogen in vanadium: a first-principles investigation},
	volume = {23},
	year = {2011}}

@mastersthesis{Courtney1977,
  author  = {D. R. Courtney},
  title   = {Thermal conductivity of hydrogen doped high purity vanadium},
  school  = {Iowa State University},
  address = {Ames, Iowa, USA},
  year    = {1977},
  type    = {M.{S}. thesis}}

@article{Li_InorgChem,
	author = {X. Li and F. Peng},
	journal = {Inorg.\ Chem.},
	pages = {13759},
	title = {Superconductivity of Pressure-Stabilized Vanadium Hydrides},
	volume = {56},
	year = {2017}}

@article{Pan_IJHE,
	author = {Y. Pan and E. Yu},
	journal = {Int.\ J.\ Hydrogen\ Energy},
	pages = {27608},
	title = {Theoretical prediction of structure, electronic and optical properties of {V}{H}$_{2}$ hydrogen storage material},
	volume = {47},
	year = {2022}}

\end{document}